\documentclass[pdftex,twocolumn,epjc3_preprint,runningheads]{svjour3}
\makeatletter
\expandafter\let\csname opt@amsmath.sty\endcsname\relax
\makeatother

\usepackage{graphicx}
\usepackage{xspace}
\usepackage{amssymb}
\usepackage{amsmath}
\usepackage{bm}
\usepackage{listings}
\usepackage{multirow}
\usepackage{array}
\usepackage{enumerate}
\usepackage[numbers,sort&compress]{natbib}
\usepackage[colorlinks,citecolor=blue,urlcolor=blue,linkcolor=blue,breaklinks=breakall]{hyperref}
\usepackage{breakurl}
\usepackage[dvipsnames]{xcolor}
\usepackage[clockwise,figuresright]{rotating}

\usepackage{graphicx,tikz,pgflibraryshapes}
\usetikzlibrary{decorations,trees}
\usetikzlibrary{positioning,shapes}
\usetikzlibrary{decorations.text}
\tikzstyle{every picture}+=[remember picture]
\pgfdeclarelayer{background}
\pgfdeclarelayer{foreground}
\pgfsetlayers{background,main,foreground}

\allowdisplaybreaks
\makeatletter
\newcommand{\preprintnumber}[1]{\gdef\@preprintnumber{\begin{flushright}{#1}\end{flushright}}}

\g@addto@macro\bfseries{\boldmath}
\makeatother

\newcommand{\hammer}{\texttt{Hammer}\xspace}
\newcommand{\PSp}{\mathcal{PS}}
\newcommand{\dds}{D^{(*)}}

\newcommand{\cbar}{\bar{c}}
\newcommand{\mn}{{\mu\nu}}

\journalname{Eur.\ Phys.\ J.\ C}
\smartqed
\makeatletter
\let\orgdescriptionlabel\descriptionlabel
\renewcommand*{\descriptionlabel}[1]{%
  \let\orglabel\label
  \let\label\@gobble
  \phantomsection
  \protected@edef\@currentlabel{#1}%
  \let\label\orglabel
  \orgdescriptionlabel{#1}%
}
\makeatother

\newcommand\cpp[1]{{\lstinline!#1!}}  

\newcommand\yaml[1]{{\lstset{style=yaml}\lstinline!#1!\lstset{style=cpp}}}

\newcommand\term[1]{{\lstset{style=terminal}\lstinline!#1!\lstset{style=cpp}}}
\newcommand\fortran[1]{{\lstset{style=fortran}\lstinline!#1!\lstset{style=cpp}}}
\newcommand\py[1]{{\lstset{style=python}\lstinline!#1!\lstset{style=cpp}}}
\newcommand\customtilde{{\raisebox{0.2ex}{\scalebox{0.6}{\boldmath$\sim$}}}}
\newcommand\mathematica[1]{{\lstset{style=Mathematica}\lstinline!#1!\lstset{style=cpp}}}

\lstnewenvironment{lstlistingyaml}{\lstset{style=yaml}}{\lstset{style=cpp}}
\lstnewenvironment{lstlistingterm}{\lstset{style=terminal}}{\lstset{style=cpp}}
\lstnewenvironment{lstlistingfortran}{\lstset{style=fortran}}{\lstset{style=cpp}}
\lstnewenvironment{lstcpp}{\lstset{style=cpp}}{\lstset{style=cpp}}
\lstnewenvironment{lstcppalt}{\lstset{style=cppalt}}{\lstset{style=cpp}}
\lstnewenvironment{lstcppnum}{\lstset{style=cppnum}}{\lstset{style=cpp}}
\lstnewenvironment{lstyaml}{\lstset{style=yaml}}{\lstset{style=cpp}}
\lstnewenvironment{lstterm}{\lstset{style=terminal}}{\lstset{style=cpp}}
\lstnewenvironment{lsttermalt}{\lstset{style=terminalalt}}{\lstset{style=cpp}}
\lstnewenvironment{lsttext}{\lstset{style=text}}{\lstset{style=cpp}}
\lstnewenvironment{lstfortran}{\lstset{style=fortran}}{\lstset{style=cpp}}
\lstnewenvironment{lstpy}{\lstset{style=python}}{\lstset{style=cpp}}
\lstnewenvironment{lstmathematica}{\lstset{style=mathematica}}{\lstset{style=cpp}}

\newcommand{\tmpname}{}
\newcommand{\tmplistingname}{}
\makeatletter
\newif\ifATOlabelname
\lst@Key{labelname}{Listing}{\def\ATOlabelname{#1}\global\ATOlabelnametrue}
\makeatother
\lstnewenvironment{lstcpplabel}[1][]{
  \lstset{style=cpp,#1} 
  \ifATOlabelname
    \renewcommand{\tmpname}{\lstlistingname}
    \renewcommand{\tmplistingname}{\lstlistlistingname}
    \renewcommand{\lstlistingname}{\ATOlabelname}
    \renewcommand{\lstlistlistingname}{List of \lstlistingname s}
  \fi
}{
  \renewcommand{\lstlistingname}{\tmpname}
  \renewcommand{\lstlistlistingname}{\tmplistingname}
  \lstset{style=cpp}
}
\definecolor{solarized@base03}{HTML}{002B36}
\definecolor{solarized@base02}{HTML}{073642}
\definecolor{solarized@base01}{HTML}{586e75}
\definecolor{solarized@base00}{HTML}{657b83}
\definecolor{solarized@base0}{HTML}{839496}
\definecolor{solarized@base1}{HTML}{93a1a1}
\definecolor{solarized@base2}{HTML}{EEE8D5}
\definecolor{solarized@base3}{HTML}{FDF6E3}
\definecolor{solarized@yellow}{HTML}{B58900}
\definecolor{solarized@orange}{HTML}{CB4B16}
\definecolor{solarized@red}{HTML}{DC322F}
\definecolor{solarized@magenta}{HTML}{D33682}
\definecolor{solarized@violet}{HTML}{6C71C4}
\definecolor{solarized@blue}{HTML}{268BD2}
\definecolor{solarized@cyan}{HTML}{2AA198}
\definecolor{solarized@green}{HTML}{859900}
\definecolor{darkred}{HTML}{550003}
\definecolor{darkgreen}{HTML}{00AA00}

\newcommand\YAMLstringstyle{\footnotesize\color{solarized@green}\mdseries}
\newcommand\YAMLkeystyle{\footnotesize\color{solarized@blue}\ttfamily}
\newcommand\YAMLvaluestyle{\footnotesize\color{blue}\mdseries}
\newcommand\ProcessThreeDashes{\llap{\color{cyan}\mdseries-{-}-}}

\newcommand\CPPcommentstyle{\color{solarized@violet}\footnotesize\ttfamily}
\newcommand\CPPdirectivestyle{\color{solarized@magenta}\footnotesize\ttfamily}
\newcommand\termplainstyle{\footnotesize\ttfamily}

\newcommand\processLongMacroDelimiter
{%
\CPPdirectivestyle%
\#define%
}

\lstdefinestyle{cpp}
{
  language=C++,
  basicstyle=\footnotesize\ttfamily,
  basewidth={0.53em,0.44em}, 
  numbers=none,
  tabsize=2,
  breaklines=true,
  escapeinside={@}{@},
  showstringspaces=false,
  numberstyle=\tiny\color{solarized@base01},
  keywordstyle=\color{solarized@orange},
  stringstyle=\color{solarized@red}\ttfamily,
  identifierstyle=\color{solarized@blue},
  commentstyle=\CPPcommentstyle,
  directivestyle=\CPPdirectivestyle,
  emphstyle=\color{solarized@green},
  frame=single,
  rulecolor=\color{solarized@base2},
  rulesepcolor=\color{solarized@base2},
  literate={~} {\customtilde}1,
  moredelim=*[directive]\ \ \#,
  moredelim=*[directive]\ \ \ \ \#,
  }

\lstdefinestyle{cppalt}
{
  language=C++,
  basicstyle=\footnotesize\ttfamily,
  basewidth={0.53em,0.44em}, 
  numbers=none,
  tabsize=2,
  breaklines=true,
  escapeinside={*@}{@*},
  showstringspaces=false,
  numberstyle=\tiny\color{solarized@base01},
  keywordstyle=\color{solarized@orange},
  stringstyle=\color{solarized@red}\ttfamily,
  identifierstyle=\color{solarized@blue},
  commentstyle=\CPPcommentstyle,
  directivestyle=\CPPdirectivestyle,
  emphstyle=\color{solarized@green},
  frame=single,
  rulecolor=\color{solarized@base2},
  rulesepcolor=\color{solarized@base2},
  literate={~}{\customtilde}1,
  moredelim=**[is][\processLongMacroDelimiter]{BeginLongMacro}{EndLongMacro} 
}

\lstdefinestyle{cppnum}
{
  language=C++,
  basicstyle=\footnotesize\ttfamily,
  basewidth={0.53em,0.44em}, 
  numbers=none,
  tabsize=2,
  breaklines=true,
  escapeinside={@}{@},
  numberstyle=\tiny\color{solarized@base01},
  showstringspaces=false,
  numberstyle=\tiny\color{solarized@base01},
  keywordstyle=\color{solarized@orange},
  stringstyle=\color{solarized@red}\ttfamily,
  identifierstyle=\color{solarized@blue},
  commentstyle=\CPPcommentstyle,
  directivestyle=\CPPdirectivestyle,
  emphstyle=\color{solarized@green},
  frame=single,
  rulecolor=\color{solarized@base2},
  rulesepcolor=\color{solarized@base2},
  literate={~} {\customtilde}1,
  moredelim=*[directive]\ \ \#,
  moredelim=*[directive]\ \ \ \ \#
}

\lstdefinestyle{python}
{
  language=Python,
  basicstyle=\footnotesize\ttfamily,
  basewidth={0.53em,0.44em},
  numbers=none,
  tabsize=2,
  breaklines=true,
  escapeinside={@}{@},
  showstringspaces=false,
  numberstyle=\tiny\color{solarized@base01},
  keywordstyle=\color{blue},
  stringstyle=\color{orange}\ttfamily,
  identifierstyle=\color{darkred},
  commentstyle=\color{purple},
  emphstyle=\color{green},
  frame=single,
  rulecolor=\color{solarized@base2},
  rulesepcolor=\color{solarized@base2},
  literate = {~}{\customtilde}1
             {\ as\ }{{\color{blue}\ as\ \color{black}}}3
}

\lstdefinestyle{fortran}
{
  language=Fortran,
  basicstyle=\footnotesize\ttfamily,
  basewidth={0.53em,0.44em},
  numbers=none,
  tabsize=2,
  breaklines=true,
  escapeinside={@}{@},
  showstringspaces=false,
  numberstyle=\tiny\color{solarized@base01},
  keywordstyle=\color{blue},
  stringstyle=\color{orange}\ttfamily,
  identifierstyle=\color{Periwinkle},
  commentstyle=\color{purple},
  emphstyle=\color{green},
  morekeywords={and, or, true, false},
  frame=single,
  rulecolor=\color{solarized@base2},
  rulesepcolor=\color{solarized@base2},
  literate={~}{\customtilde}1
}

\lstdefinestyle{terminal}
{
  language=bash,
  basicstyle=\termplainstyle,
  numbers=none,
  tabsize=2,
  breaklines=true,
  escapeinside={@}{@},
  frame=single,
  showstringspaces=false,
  numberstyle=\tiny\color{solarized@base01},
  keywordstyle=\color{solarized@orange},
  stringstyle=\color{solarized@red}\ttfamily,
  identifierstyle=\color{black},
  commentstyle=\color{solarized@violet},
  emphstyle=\color{solarized@green},
  frame=single,
  rulecolor=\color{solarized@base2},
  rulesepcolor=\color{solarized@base2},
  morekeywords={gambit, cmake, make, mkdir},
  deletekeywords={test},
  literate = {\ gambit}{{\ }{\color{black}}gambit}7
             {/gambit}{{/}{\color{black}}gambit}6
             {gambit/}{{\color{black}}gambit{/}}6
             {/include}{{/}{\color{black}}include}8
             {cmake/}{{\color{black}}cmake/}6
             {.cmake}{{.}{\color{black}}cmake}6
             {~}{\customtilde}1
}

\lstdefinestyle{terminalalt}
{
  language=bash,
  basicstyle=\footnotesize\ttfamily,
  numbers=none,
  tabsize=2,
  breaklines=true,
  escapeinside={*@}{@*},
  frame=single,
  showstringspaces=false,
  numberstyle=\tiny\color{solarized@base01},
  keywordstyle=\color{solarized@orange},
  stringstyle=\color{solarized@red}\ttfamily,
  identifierstyle=\color{black},
  commentstyle=\color{solarized@violet},
  emphstyle=\color{solarized@green},
  frame=single,
  rulecolor=\color{solarized@base2},
  rulesepcolor=\color{solarized@base2},
  morekeywords={gambit, cmake, make, mkdir},
  deletekeywords={test},
  literate = {\ gambit}{{\ }{\color{black}}gambit}7
             {/gambit}{{/}{\color{black}}gambit}6
             {gambit/}{{\color{black}}gambit{/}}6
             {/include}{{/}{\color{black}}include}8
             {cmake/}{{\color{black}}cmake/}6
             {.cmake}{{.}{\color{black}}cmake}6
             {~}{\customtilde}1
}

\lstdefinestyle{text}
{
  language={},
  basicstyle=\footnotesize\ttfamily,
  identifierstyle=\color{black},
  numbers=none,
  tabsize=2,
  breaklines=true,
  escapeinside={*@}{@*},
  showstringspaces=false,
  frame=single,
  rulecolor=\color{solarized@base2},
  rulesepcolor=\color{solarized@base2},
  literate={~}{\customtilde}1
}

\lstdefinestyle{yaml}
{
  language=bash,
  escapeinside={@}{@},
  keywords={true,false,null},
  otherkeywords={},
  keywordstyle=\color{solarized@base0}\bfseries,
  basicstyle=\footnotesize\color{black}\ttfamily,
  identifierstyle=\YAMLkeystyle,
  sensitive=false,
  commentstyle=\color{solarized@orange}\ttfamily,
  morecomment=[l]{\#},
  morecomment=[s]{/*}{*/},
  stringstyle=\YAMLstringstyle\ttfamily,
  moredelim=**[s][\YAMLkeystyle]{,}{:},   
  moredelim=**[l][\YAMLvaluestyle]{:},    
  morestring=[b]',
  morestring=[b]",
  literate =    {---}{{\ProcessThreeDashes}}3
                {>}{{\textcolor{solarized@red}\textgreater}}1
                {|}{{\textcolor{solarized@red}\textbar}}1
                {\ -\ }{{\mdseries\color{black}\ -\ \negmedspace}}3
                {\}}{{{\color{black} \}}}}1
                {\{}{{{\color{black} \{}}}1
                {[}{{{\color{black} [}}}1
                {]}{{{\color{black} ]}}}1
                {~}{\customtilde}1,
  breakindent=0pt,
  breakatwhitespace,
  columns=fullflexible
}

\lstdefinestyle{mathematica}
{
  language={Mathematica},
  basicstyle=\footnotesize\ttfamily,
  basewidth={0.53em,0.44em},
  numbers=none,
  tabsize=2,
  breaklines=true,
  escapeinside={@}{@},
  numberstyle=\tiny\color{black},
  showstringspaces=false,
  numberstyle=\tiny\color{solarized@base01},
  keywordstyle=\color{solarized@orange},
  stringstyle=\color{solarized@red}\ttfamily,
  identifierstyle=\color{solarized@orange}\ttfamily,
  commentstyle=\color{solarized@gray}\ttfamily,
  directivestyle=\color{solarized@orange}\ttfamily,
  emphstyle=\color{solarized@green},
  frame=single,
  rulecolor=\color{solarized@base2},
  rulesepcolor=\color{solarized@base2},
  literate={~} {\customtilde}1,
  moredelim=*[directive]\ \ \#,
  moredelim=*[directive]\ \ \ \ \#,
  mathescape=true
}

\newcommand\textlst[1]{\lstinline!#1!}

\newcommand{\bcode}{\begin{lstlisting}}
\newcommand{\ecode}{\end{lstlisting}}

\newcommand\beq{\begin{equation}}
\newcommand\eeq{\end{equation}}

\renewcommand{\url}[1]{\href{#1}{#1}}

\def\MagUp {\mbox{\em Mag\kern -0.05em Up}\xspace}

 \def\PDelta      {\ensuremath{\Delta}\xspace}
 \def\PXi      {\ensuremath{\Xi}\xspace}
 \def\PLambda      {\ensuremath{\Lambda}\xspace}
 \def\PSigma      {\ensuremath{\Sigma}\xspace}
 \def\POmega      {\ensuremath{\Omega}\xspace}
 \def\PUpsilon      {\ensuremath{\Upsilon}\xspace}

 \def\PB      {\ensuremath{\mathrm{B}}\xspace}
 
 \def\PD      {\ensuremath{\mathrm{D}}\xspace}

 \def\PK      {\ensuremath{\mathrm{K}}\xspace}

 \def\Pi      {\ensuremath{\mathrm{i}}\xspace}

{

 \mathchardef\PDelta="7101
 \mathchardef\PXi="7104
 \mathchardef\PLambda="7103
 \mathchardef\PSigma="7106
 \mathchardef\POmega="710A
 \mathchardef\PUpsilon="7107
 
 \def\PB      {\ensuremath{B}\xspace}
 
 \def\PD      {\ensuremath{D}\xspace}

 \def\PK      {\ensuremath{K}\xspace}

 \def\Pi      {\ensuremath{i}\xspace}

}

\makeatletter
\ifcase \@ptsize \relax
  \newcommand{\miniscule}{\@setfontsize\miniscule{4}{5}}
\or
  \newcommand{\miniscule}{\@setfontsize\miniscule{5}{6}}
\or
  \newcommand{\miniscule}{\@setfontsize\miniscule{5}{6}}
\fi
\makeatother

\DeclareRobustCommand{\optbar}[1]{\shortstack{{\miniscule (\rule[.5ex]{1.25em}{.18mm})}
  \\ [-.7ex] $#1$}}

  \def\Kbar    {{\kern 0.2em\overline{\kern -0.2em \PK}{}}\xspace}

\def\KorKbar    {\kern 0.18em\optbar{\kern -0.18em K}{}\xspace}

  \def\Dbar    {{\kern 0.2em\overline{\kern -0.2em \PD}{}}\xspace}

\def\DorDbar    {\kern 0.18em\optbar{\kern -0.18em D}{}\xspace}

\def\Bbar    {{\ensuremath{\kern 0.18em\overline{\kern -0.18em \PB}{}}}\xspace}

\def\BorBbar    {\kern 0.18em\optbar{\kern -0.18em B}{}\xspace}

  \def\Y#1S{\ensuremath{\PUpsilon{(#1S)}}\xspace}

\def\Lbar        {{\ensuremath{\kern 0.1em\overline{\kern -0.1em\PLambda}}}\xspace}
\def\LorLbar    {\kern 0.18em\optbar{\kern -0.18em \PLambda}{}\xspace}

\def\BF         {{\ensuremath{\cal B}}\xspace}

\def\BR         {\BF}

\def\to                 {\ensuremath{\rightarrow}\xspace}

\newcommand{\lqcd}{{\ensuremath{\Lambda_{\mathrm{QCD}}}}\xspace}

\def\AT#1     {\ensuremath{A_{\mathrm{T}}^{#1}}\xspace}           

\def\C#1      {\ensuremath{\mathcal{C}_{#1}}\xspace}                       
\def\Cp#1     {\ensuremath{\mathcal{C}_{#1}^{'}}\xspace}                    
\def\Ceff#1   {\ensuremath{\mathcal{C}_{#1}^{\mathrm{(eff)}}}\xspace}        
\def\Cpeff#1  {\ensuremath{\mathcal{C}_{#1}^{'\mathrm{(eff)}}}\xspace}       
\def\Ope#1    {\ensuremath{\mathcal{O}_{#1}}\xspace}                       
\def\Opep#1   {\ensuremath{\mathcal{O}_{#1}^{'}}\xspace}                    

\newcommand{\tev}{\ifthenelse{\boolean{inbibliography}}{\ensuremath{~T\kern -0.05em eV}\xspace}{\ensuremath{\mathrm{\,Te\kern -0.1em V}}}\xspace}
\newcommand{\gev}{\ensuremath{\mathrm{\,Ge\kern -0.1em V}}\xspace}
\newcommand{\mev}{\ensuremath{\mathrm{\,Me\kern -0.1em V}}\xspace}
\newcommand{\kev}{\ensuremath{\mathrm{\,ke\kern -0.1em V}}\xspace}
\newcommand{\ev}{\ensuremath{\mathrm{\,e\kern -0.1em V}}\xspace}
\newcommand{\gevc}{\ensuremath{{\mathrm{\,Ge\kern -0.1em V\!/}c}}\xspace}
\newcommand{\mevc}{\ensuremath{{\mathrm{\,Me\kern -0.1em V\!/}c}}\xspace}
\newcommand{\gevcc}{\ensuremath{{\mathrm{\,Ge\kern -0.1em V\!/}c^2}}\xspace}
\newcommand{\gevgevcccc}{\ensuremath{{\mathrm{\,Ge\kern -0.1em V^2\!/}c^4}}\xspace}
\newcommand{\mevcc}{\ensuremath{{\mathrm{\,Me\kern -0.1em V\!/}c^2}}\xspace}

\def\gsim{{~\raise.15em\hbox{$>$}\kern-.85em
          \lower.35em\hbox{$\sim$}~}\xspace}
\def\lsim{{~\raise.15em\hbox{$<$}\kern-.85em
          \lower.35em\hbox{$\sim$}~}\xspace}

\def\fortran    {\mbox{\textsc{Fortran}}\xspace}

\def\tell1  {TELL1\xspace}
\def\ukl1   {UKL1\xspace}

\preprintnumber{}

\title{Das ist der HAMMER: Consistent new physics interpretations of semileptonic decays}

\newcommand{\bonn}{Physikalisches Institut der Rheinischen Friedrich-Wilhelms-Universit\"at Bonn, 53115 Bonn, Germany}
\newcommand{\lbnl}{Ernest Orlando Lawrence Berkeley National Laboratory, University of California, Berkeley, CA 94720, USA}
\newcommand{\caltech}{Burke Institute for Theoretical Physics, California Institute of Technology, Pasadena, CA 91125, USA}

\institute{%
  \bonn\label{inst:b} \and
  \lbnl\label{inst:l} \and
  \caltech\label{inst:c}     
}

\thankstext{e1}{florian.bernlochner@uni-bonn.de}
\thankstext{e2}{s.duell@physik.uni-bonn.de}
\thankstext{e3}{ligeti@berkeley.edu}
\thankstext{e4}{mpapucci@caltech.edu}
\thankstext{e5}{drobinson@lbl.gov}

\author{
Florian U.\ Bernlochner\thanksref{e1,inst:b} \and
Stephan Duell\thanksref{e2,inst:b} \and
Zoltan Ligeti\thanksref{e3,inst:l} \and
\hspace*{.5cm}\raisebox{-16pt}[0pt][0pt]{\includegraphics[width=1.2cm]{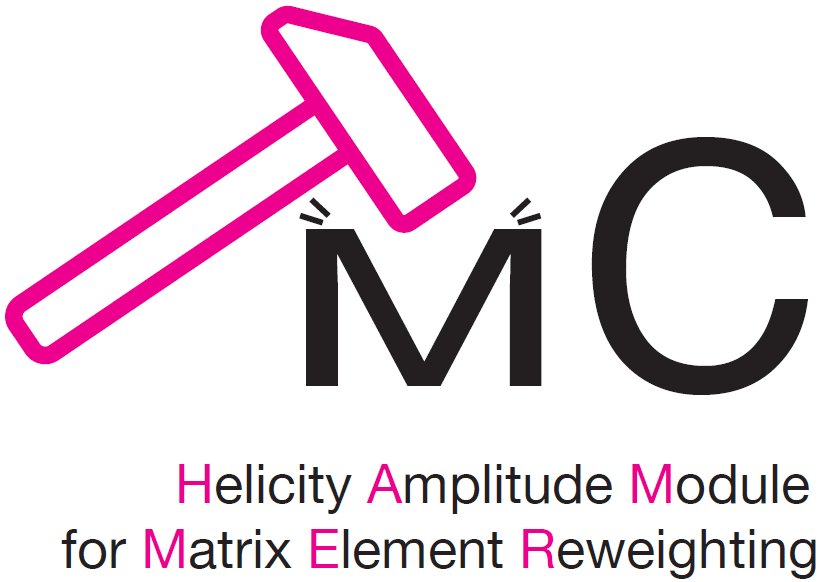}}
Michele~Papucci\thanksref{e4,inst:l,inst:c} \and
Dean J.\ Robinson\thanksref{e5,inst:l}
}

\date{Received: ... / Accepted: ...}
\date{}

\begin{document}

\maketitle

\begin{abstract}
Precise measurements of $b\to c\tau\bar\nu$ decays require large resource-intensive Monte Carlo (MC) samples, 
which incorporate detailed simulations of detector responses and physics backgrounds. 
Extracted parameters may be highly sensitive to the underlying theoretical models used in the MC generation.
Because new physics (NP) can alter decay distributions and acceptances,
the standard practice of fitting NP Wilson coefficients to SM-based measurements of the $R(\dds)$ ratios  
can be biased. 
The newly developed \hammer software tool enables efficient reweighting of MC samples to arbitrary NP scenarios or to any hadronic matrix elements. 
We demonstrate how \hammer allows avoidance of biases through self-consistent fits directly to the NP Wilson coefficients. 
We also present example analyses that demonstrate the sizeable biases that can otherwise occur from naive NP interpretations of SM-based measurements. 
The \hammer library is presently interfaced with several existing experimental analysis frameworks and we provide an overview of its structure.
\end{abstract}

\tableofcontents

\section{Introduction}
\label{sec:introduction}

Precision analyses of semileptonic $b$-hadron decays typically rely on detailed numerical Monte Carlo (MC) simulations of detector responses and acceptances.
Combined with the underlying theoretical models, these simulations provide MC \emph{templates} that may be used in fits, to
translate experimental yields into theoretically well-defined parameters.
This translation though can become sensitive to the template and its underlying theoretical model, 
introducing biases whenever there is a mismatch between the theoretical assumptions used to measure a parameter and subsequent theoretical interpretations of the data.

Such biases are known to arise in the analyses of semileptonic decays of $b$ hadrons, in particular, for the measurements of the CKM element $|V_{cb}|$, and in the ratio of semitauonic vs.\ semileptonic decays to light leptons,~(see e.g.~Refs.~\cite{Lees:2013uzd, Huschle:2015rga} and Ref.~\cite{Tanabashi:2018oca}, respectively),
\begin{equation}
\label{RMdef}
	R(H_c) = \frac{\Gamma(H_b\to H_c\tau\bar\nu)}{\Gamma(H_b\to H_c l\bar\nu)}\,, 
  \qquad l = \mu, \,e\,,
\end{equation}
where $H_{b,c}$ denote $b$-~and $c$-flavor hadrons.
To avoid this, the size of these biases need to be either carefully controlled when experiments quote their results by reversing detector effects, 
or they can be avoided by using dedicated MC samples for each theoretical model the measurement is confronted with. 
In this paper we present the newly developed tool, \hammer\ (\emph{Helicity Amplitude Module for Matrix Element Reweighting}), designed expressly for the latter purpose.

Semitauonic $b$ hadron decays have long been known to be sensitive to new physics~\cite{Krawczyk:1987zj, 
Heiliger:1989yp, Kalinowski:1990ba, Grzadkowski:1991kb, Grossman:1994ax, Tanaka:1994ay, Goldberger:1999yh}, 
and were first constrained at LEP~\cite{Buskulic:1992um}. 
At present, the measurements of the $R(\dds)$ ratios show about a $3\sigma$ tension with SM predictions, when the $D$ and $D^*$ modes are combined~\cite{Amhis:2019ckw}. 
In the future, much more precise measurements of semitauonic decays are expected, not only for the $B\to D^{(*)}\tau\bar\nu$ channels, 
but also for the not yet studied decay modes, $\Lambda_b\to \Lambda_c\tau\bar\nu$, $B_s\to D_s^{(*)}\tau\bar\nu$, as well as involving excited charm hadrons in the final state.

All existing measurements of $R(\dds)$ rely heavily on large MC simulations to optimize selections, provide fit templates in discriminating kinematic observables, 
and to model resolution effects and  acceptances. Both the $\tau$ and the charm hadrons have short lifetimes and decay near the interaction point and measurements
rely on reconstruction of the ensuing decay cascades. To reconstruct the decay products, often complex phase space cuts and detector efficiency dependencies come into play, 
and the measurement of the full decay kinematics is impossible due to the presence of multiple neutrinos.
In addition, depending on the final state, a significant downfeed with similar experimental signatures from misreconstructed excited charm hadron states can be present. 
Isolation of semitauonic decays from other background processes and the light-lepton final states, 
then requires precise predictions for the kinematics of the signal semitauonic decay.\footnote{
Further complications arise from interference among the different spin states of the $\tau$ and among those of the charm hadron. 
Such effects have sometimes been neglected, treating the $\tau$ and charm hadron as stable particles,
when simulations are corrected to account for more up-to-date hadronic models.} 
Often the limited size of the available simulated samples, required to account for all these effects, constitutes a dominant uncertainty of the measurements, see e.g.~\cite{Lees:2013uzd,Aaij:2015yra,Huschle:2015rga}.

In the literature on the $R(\dds)$ anomaly, 
it has become standard practice to reinterpret the experimental values of $R(\dds)$ in terms of NP Wilson coefficients, 
even though all current ratio measurements were determined assuming the SM nature of semitauonic decays. 
However, NP couplings generically alter decay distributions and acceptances. 
Therefore, they modify the signal and possibly background MC templates used in the extraction, and thus affect the measured values of $R(\dds)$.  
This may introduce biases in NP interpretations: preferred regions and best-fit points for the Wilson coefficients can be incorrect; an instructive example of this is provided in Sec~\ref{subsec:NP_bias}.  

Consistent interpretations of the data with NP incorporated requires dedicated MC samples, ideally for each NP coupling value considered, 
which would permit directly fitting for the NP Wilson coefficients.
This approach is sometimes referred to as `forward-folding', and is naively a computationally prohibitively expensive endeavour. Such a program is further complicated because none of the MC generators current used by the experiments incorporate generic NP effects, nor do they include state-of-the-art treatments of hadronic matrix elements.

In this paper we present a new software tool, \hammer, that provides a solution to these problems: 
A fast and efficient means to reweight large MC samples to any desired NP, or to any description of the hadronic matrix elements.
 \hammer makes use of efficient amplitude-level and tensorial calculation strategies,
and is designed to interface with existing experimental analysis frameworks,
providing detailed control over which NP or hadronic descriptions should be considered. 
The desired reweighting can be implemented either in the event weights or 
in histograms of experimentally reconstructed quantities (both further discussed in Sec.~\ref{sec:code}). 
The only required input are the event-level truth-four-momenta of existing MC samples. 
Either the event weights and/or histogram predictions may be used, e.g., to generate likelihood functions for experimental fits. 
Some of the main ideas of \hammer were previously outlined in Refs.~\cite{Ligeti:2016npd, Duell:2016maj}. 

In Sec.~\ref{sec:NP_ana} we demonstrate the capabilities of \hammer by performing binned likelihood fits on mock measured and simulated data sets, 
that are created using the  \hammer\ library, and corrected using an approximate detector response. 
In Sec.~\ref{sec:code} a brief overview of the \hammer\ library and its capabilities are given. Section~\ref{sec:summary} provides a summary of our findings. 
Finally, Appendix~\ref{sec:API} provides a detailed overview of the \hammer\ application programming interface. 

\section{New physics analyses}
\label{sec:NP_ana}

We consider two different analysis scenarios:
\begin{enumerate}
\item In order to explore what biases may arise in phenomenological studies if NP is present in Nature, we perform an illustrative $R(\dds)$ toy measurement. 
This involves carrying out SM fits to mockups of measured data sets, that are generated for several different NP models. 
The recovered $R(\dds)$ values are then compared to their actual NP values. 
\item To demonstrate using a forward-folded analysis to assess NP effects without biases, we carry out fits to (combinations of) NP Wilson coefficients themselves, 
with either the SM or other NP present in the mock measured data sets.
\end{enumerate}
The setting of these analyses is a $B$-factory-type environment. 
We focus on leptonic $\tau$ decays, but the procedures and results in this work are equally adaptable to the LHCb environment, and other $\tau$ decay modes or observables. In our example we focus on kinematic observables important for the separation of signal from background and normalization modes. Fits using angular information may also be implemented, see e.g. Refs.~\cite{Hill:2019zja,Becirevic:2019tpx} for an example.

We emphasize that the derived sensitivities shown below are not intended to illustrate projections for actual experimental sensitivities \emph{per se}.
Such studies are better carried out by the experiments themselves. 

\subsection{MC sample}

The input Monte Carlo sample used for our demonstration comprises four distinct sets of $10^5$ events: one for each of the two signal cascades 
$B \to D (\tau \to e\nu\nu)\nu$, $B \to (D^* \to D\pi) (\tau \to e\nu\nu) \nu$ and for the two background processes, $B \to D e\nu$ and $B \to (D^* \to D\pi) e\nu$. 
These are generated with \texttt{EvtGen R01-07-00}~\cite{Lange:2001uf}, using the Belle~II beam energies of~$7$\,GeV and $4$\,GeV. 
The second $B$ meson decay, often used for identifying or `tagging' the $B\bar B$ event and constraining its kinematic properties, 
are not included in the current analysis for simplicity, but can be incorporated in a \hammer analysis straightforwardly. 

In each cascade, the $b \to c l \nu$ decay is generated equidistributed in phase space (``pure phase space''), instead of using SM distributions.   
This reduces the statistical uncertainties that can otherwise arise from reweighting regions of phase space that are undersampled in the SM to NP scenarios 
in which they are not.\footnote{
For an actual experimental analysis one would instead use \hammer to reweight SM MC samples.
The correct statistical uncertainty of the reweighting can be incorporated, using weight squared uncertainties computed by the library. 
This information could be used, e.g., to adaptively generate additional pure phase space MC in undersampled regions.}

\subsection{Reweighting and fitting analysis}
\label{sec:rwgt}

\hammer is used to reweight the MC samples into two-dimensional `NP generalized' histograms (see Sec.~\ref{sec:code}), 
with respect to the reconstructed observables $| \vec p^*_{\ell}|$ and $m^2_{\text{miss}}$, 
the light lepton momentum in the $B$ rest frame and the total missing invariant mass of all neutrinos, respectively. 
Both variables are well-suited for separating signal from background decays involving only light leptons. 
In the cascade process of the leptonic $\tau$ decay in $B \to D^{(*)} \tau \nu$, the signal lepton carries less momentum than the lepton from prompt $B \to D^{(*)} \ell \nu$ decays. 
Similarly, the missing invariant mass of  $B \to D^{(*)} \ell \nu$ decays peaks strongly near $m_\nu^2 \simeq 0$, 
in contrast to $B \to D^{(*)} \tau \nu$ in which the multiple neutrinos in the final state permit large values of $m^2_{\text{miss}}$. 

The $B \to \dds$ processes are reweighted to the BLPR form factor parametrization~\cite{Bernlochner:2017jka}, 
which includes predictions for NP hadronic matrix elements using HQET~\cite{Isgur:1989vq, Isgur:1989ed, Eichten:1989zv, Georgi:1990um} at $\mathcal{O}(1/m_{c,b},\, \alpha_s)$.

Charged particles are required to fall in the Belle~II angular acceptance of $20^\circ$ and $150^\circ$, 
and leptons are required to have a minimum kinetic energy of $300$\,MeV in the laboratory frame.  
An additional event weight is included to account for the slow pion reconstruction efficiencies from the $D^* \to D\pi$ decay, 
based on an approximate fit to the pion reconstruction efficiency curve from BaBar data~\cite{Lees:2012xj, Lees:2013uzd}. 
The analysis assumes that the second tagging $B$ meson decay was reconstructed in hadronic modes, such that its four-momentum, $p_{B_{\rm tag}}$, is accessible. 
In conjunction with the known beam four-momentum $p_{e^+ \, e^-}$, the missing invariant mass can then be reconstructed as 
$m^2_{\text{miss}} \equiv (p_{e^+ \, e^-} -  p_{B_{\rm tag}} - p_{\dds} - p_\ell)^2$, and the four-momentum of the reconstructed lepton can be boosted 
into the signal $B$ rest frame. 
A Gaussian smearing is added to the truth level $m^2_{\text{miss}}$ with a width of $0.5$\,GeV$^2$ to account for detector resolution and tagging-$B$ reconstruction. 
No additional correction is applied to $|\vec p^*_{\ell}|$. 
Higher dimensional histograms including the reconstructed $q^2$ and the $D^* \to D\pi$ helicity angle may also be incorporated, but are omitted here for simplicity.

\hammer\ can be used to efficiently compute histograms for any given NP choice. 
The basis of NP operators is defined in Table~\ref{tab:NPc}, with respect to the Lagrangian
\begin{equation}\label{Ltimesi}
	\mathcal{L} = \frac{4 G_F}{\sqrt 2}\, V_{cb}\,  c_{XY}\big(\cbar\, \Gamma_X\, b\big)\big(\bar\ell\, \Gamma_Y\, \nu\big)\,,
\end{equation}	
where $\Gamma_{X(Y)}$ is any Dirac matrix and $c_{XY}$ is a Wilson coefficient. We shall generally write explicit Wilson coefficients as
$c_{XY} = S_{qXlY}$, $V_{qXlY}$, $T_{qXlY}$, where the $S$, $V$, $T$ denotes the Lorentz structure, and $X$, $Y$ = $L$, $R$ denotes the chirality.
In this simplified analysis, we assume that NP only affects the $b \to c \tau \nu$ decays, and not the light-lepton modes. 

In order to carry out Wilson coefficient fits, we wrap the \hammer application programming interface with a \texttt{gammaCombo}~\cite{Aaij:2016kjh} compatible class. 
This allows one to use \hammer's efficient reweighting of histogram bins to generate the relevant quantities required to calculate a likelihood function 
for the binned observables of interest. 
We then carry out a fully two-dimensional binned likelihood fit in $|\vec p^*_{\ell}|$ and $m^2_{\text{miss}}$, assuming Gaussian uncertainties. 
The fit uses $12 \times 12$ bins with equidistant bin widths for $|\vec p^*_{\ell}| \in (0.2,\, 2.2)$\,GeV and $m^2_{\text{miss}} \in (-2,\, 10)$\,GeV${}^{2}$. 
The fits determine either $R(\dds)$, or the real and imaginary parts of Wilson coefficients. The preferred SM coupling is determined simultaneously,
in order to remove explicit dependence on $|V_{cb}|$.

\begin{table}[tb]
\renewcommand{\arraystretch}{1.5}
\newcolumntype{C}{ >{\centering\arraybackslash} m{1cm} <{}}
\def\simplecollect#1#2\ignorespaces#3\unskip{#1{#3}\unskip}
\newcommand*{\ctexttt}[1]{\centering{\textlst{#1}}}
\newcolumntype{S}{>{\simplecollect\ctexttt} m{1cm}}
\newcolumntype{D}{ >{\raggedright\arraybackslash $} c <{$}}
\scalebox{0.85}{
\begin{tabular}{C|SDD}
\hline\hline
Current 				&  Label 	& \text{Wilson Coefficient}, c_{XY}		& \text{Operator}  \\
\hline
 SM 					& SM		& 	1			&\big[\cbar \gamma^\mu P_L b\big] \big[\bar\ell \gamma_\mu P_L \nu\big] \\
 \hline
 \multirow{4}{*}{Vector} 	& V_qLlL 		& V_{qLlL} 		&\big[\cbar \gamma^\mu P_L b\big] \big[\bar\ell \gamma_\mu P_L\nu\big] \\
 					& V_qRlL		& V_{qRlL} 		&\big[\cbar \gamma^\mu P_R b\big] \big[\bar\ell \gamma_\mu P_L\nu\big] \\
					& V_qLlR		& V_{qLlR} 		&\big[\cbar \gamma^\mu P_L b\big] \big[\bar\ell  \gamma_\mu P_R\nu\big] \\
					& V_qRlR		& V_{qRlR} 		&\big[\cbar \gamma^\mu P_R b\big] \big[\bar\ell \gamma_\mu P_R\nu\big] \\
\hline
\multirow{4}{*}{Scalar} 	& S_qLlL 		& S_{qLlL} 		&\big[\cbar  P_L b\big] \big[\bar\ell  P_L \nu\big] \\
					& S_qRlL 		& S_{qRlL}		&\big[\cbar P_R b\big] \big[\bar\ell  P_L \nu\big] \\
					& S_qLlR 		& S_{qLlR}		&\big[\cbar  P_L b\big] \big[\bar\ell P_R \nu\big] \\
					& S_qRlR 		& S_{qRlR}		&\big[\cbar  P_R b\big] \big[\bar\ell  P_R \nu\big] \\
\hline
\multirow{2}{*}{Tensor} 	& T_qLlL 		& T_{qLlL} 		&\big[\cbar \sigma^\mn P_L b\big] \big[\bar\ell \sigma_\mn P_L \nu \big] \\
					& T_qRlR 		& T_{qRlR} 		& \big[\cbar \sigma^\mn P_R b\big] \big[\bar\ell  \sigma_\mn P_R \nu \big] \\
\hline\hline
\end{tabular}
}
\caption{The $b \to c \ell \nu$ operator basis and coupling conventions. Also shown are the identifying Wilson coefficient labels used in \hammer.  The normalization of the operators is as in Eq.~(\ref{Ltimesi}).}
\label{tab:NPc}
\end{table}

We construct an Asimov data set~\cite{Cowan:2010js} assuming the fractions and total number of events in Table~\ref{tab:asimov}, 
following from the number of events in Ref.~\cite{Lees:2012xj, Lees:2013uzd}.
In the scans, the total number of events corresponds to an approximate integrated luminosity of $5 \, \text{ab}^{-1}$ of Belle~II collisions. 
We assume events are reconstructed in two categories targeting $B \to D \, \tau \bar \nu$ and $B \to D^* \tau \bar \nu$. 
A fit for the real and imaginary parts of a single Wilson coefficient plus the (real) SM coupling thus has $2 \times 12 \times 12 - 3 = 285$ degrees of freedom.  

\begin{table}[t]
\begin{center}
\newcolumntype{D}{ >{\centering\arraybackslash} m{0.4\linewidth} <{}}
\newcolumntype{E}{ >{\centering\arraybackslash} m{0.2\linewidth} <{}}
\newcolumntype{F}{ >{\centering\arraybackslash} m{0.25\linewidth} <{}}
\begin{tabular}{D|E|F}
\hline\hline
$ B \to D \tau \bar \nu$ Category &   Fractions & Events / $\text{ab}^{-1}$ \\
\hline
$ B \to D \tau \bar \nu$ & 5.6\% & 800  \\
$ B \to D^{*} \tau \bar \nu$ & 2.3\% & 325 \\
$ B \to D\ell \bar \nu$ & 49.4\% & 7000  \\
$ B \to D^{*} \ell \bar \nu$ & 40.6\% & 5750 \\
Irreducible background & 2.0\% &288 \\
\hline\hline
$ B \to D^* \tau \bar \nu$ Category &   Fractions & Events / $\text{ab}^{-1}$ \\
\hline
$ B \to D^{*} \tau \bar \nu$ & 5.4\% & 950 \\
$ B \to D^{*} \ell \bar \nu$ & 93.0\%  & 16500 \\
Irreducible background &  1.6\% &288 \\
\hline\hline
\end{tabular}
\caption{The Asimov data set components. The fractions were motivated by Refs.~\cite{Lees:2012xj, Lees:2013uzd}.}
\label{tab:asimov}
\end{center}
\end{table}

A sizable downfeed background from $D^*$ mesons misreconstructed as a $D$ is expected in the $B \to D \, \tau \bar \nu$ channel 
via both the $B \to D^* \, \tau \bar \nu$ and  $B \to D^* \, \ell \bar \nu$ decays. 
This is taken into account by partitioning the simulated $B \to D^*\tau\nu$ and $B \to D^*\ell\nu$ events into two samples: 
One with the correct $m^2_{\text{miss}} = (p_B - p_{D^*} - p_\ell)^2$ and the other with the misreconstructed $m^2_{\text{miss}} =  (p_B - p_{D} - p_\ell)^2$, which omits the slow pion. 
This downfeed reduces the sensitivity for the case that NP couplings induce opposite effects on the $B \to D \tau \bar \nu$ versus 
$B \to D^* \tau \bar \nu$ total rates or shapes.
In addition to semileptonic processes, we assume the presence of an irreducible background from secondaries (i.e., leptons from semileptonic $D$ meson decays), 
fake leptons (i.e., hadrons that were misidentified as leptons) and semileptonic decays from higher charm resonances (i.e., $D^{**}$ states).
The irreducible background is modeled in a simplified manner by assuming $10$ background events in each of the $12 \times 12$ bins, totaling overall $1440$ events per category. 

\begin{figure}[tb]
\includegraphics[width=\columnwidth]{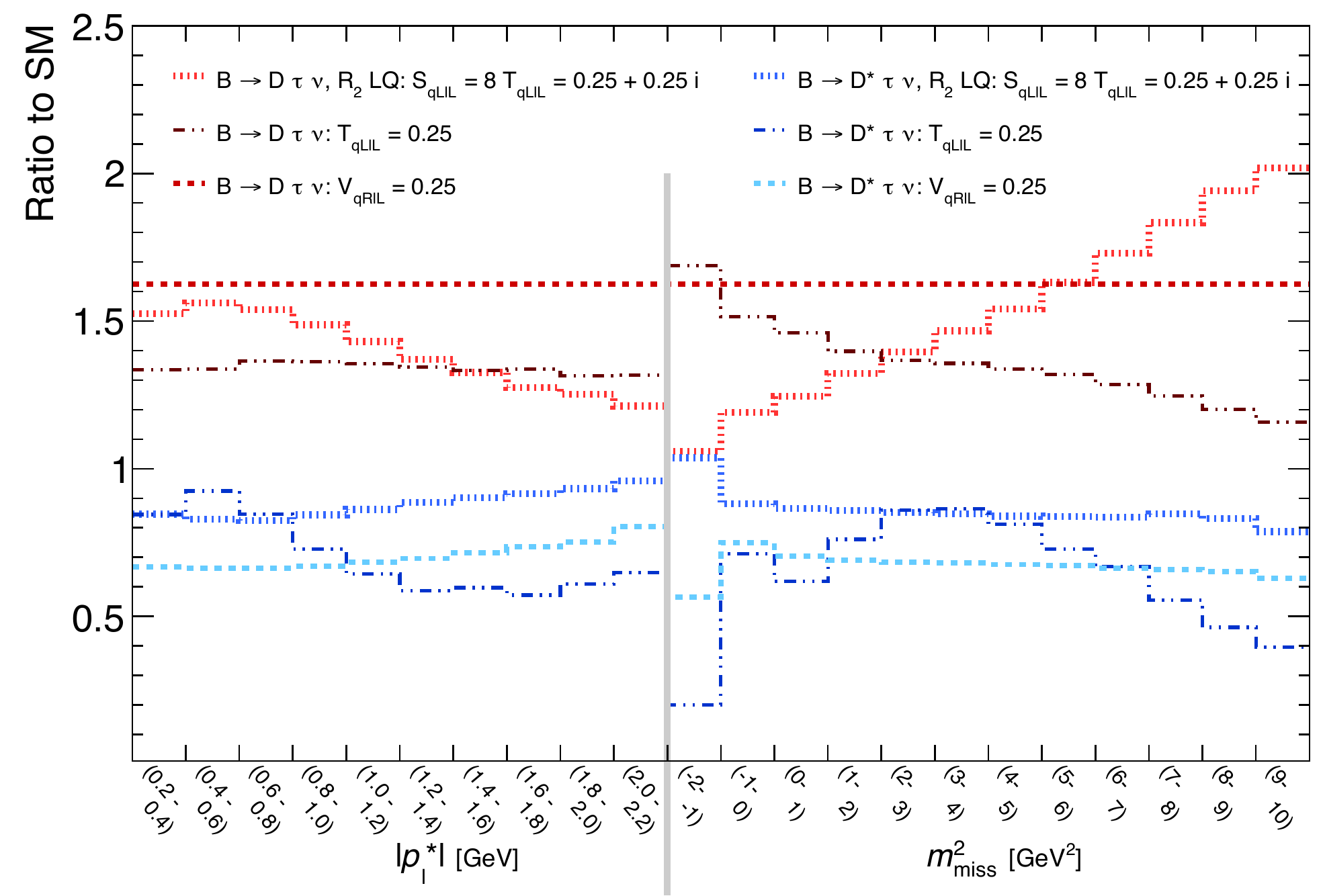}  
\caption{The ratios of differential distributions with respect to the SM, as functions of $|\vec p^*_{\ell}|$ and $m^2_{\text{miss}}$, for various Wilson coefficient working points.
For more details see text.}
\label{fig:NP_D_Ds_ratio}
\end{figure}

Figure~\ref{fig:NP_D_Ds_ratio} shows the impact on the fit variables of three benchmark models that we use to investigate the effects of new physics:
\begin{enumerate}[i)]
	\item  The $R_2$ leptoquark model, which sets $S_{qLlL} \simeq 8 \, T_{qLlL}$ (including RGE; see, e.g., Refs.~\cite{Dorsner:2016wpm, Freytsis:2015qca});
	\item A pure tensor model, via $T_{qLlL}$;
	\item A right-handed vector model, via $V_{qRlL}$\,. 
\end{enumerate}
For the ratio plots in Fig.~\ref{fig:NP_D_Ds_ratio}, we fix the NP Wilson coefficients to specific values 
to illustrate the shape changes they induce in $|\vec p^*_{\ell}|$ and $m^2_{\text{miss}}$. 
The $R_2$ leptoquark model and tensor model exhibit sizable shape changes. 
The right-handed vector model shows only an overall normalization change for $B \to D \, \tau \bar \nu$, with no change in shape compared to the SM, 
because the axial-vector $B \to D$ hadronic matrix element vanishes by parity and angular momentum conservation.
For $B \to D^*$, both vector and axial vector matrix elements are nonzero, so that introducing a right-handed vector current leads to shape and normalization changes. 

\begin{figure}[tb]
  \includegraphics[width=0.47\textwidth]{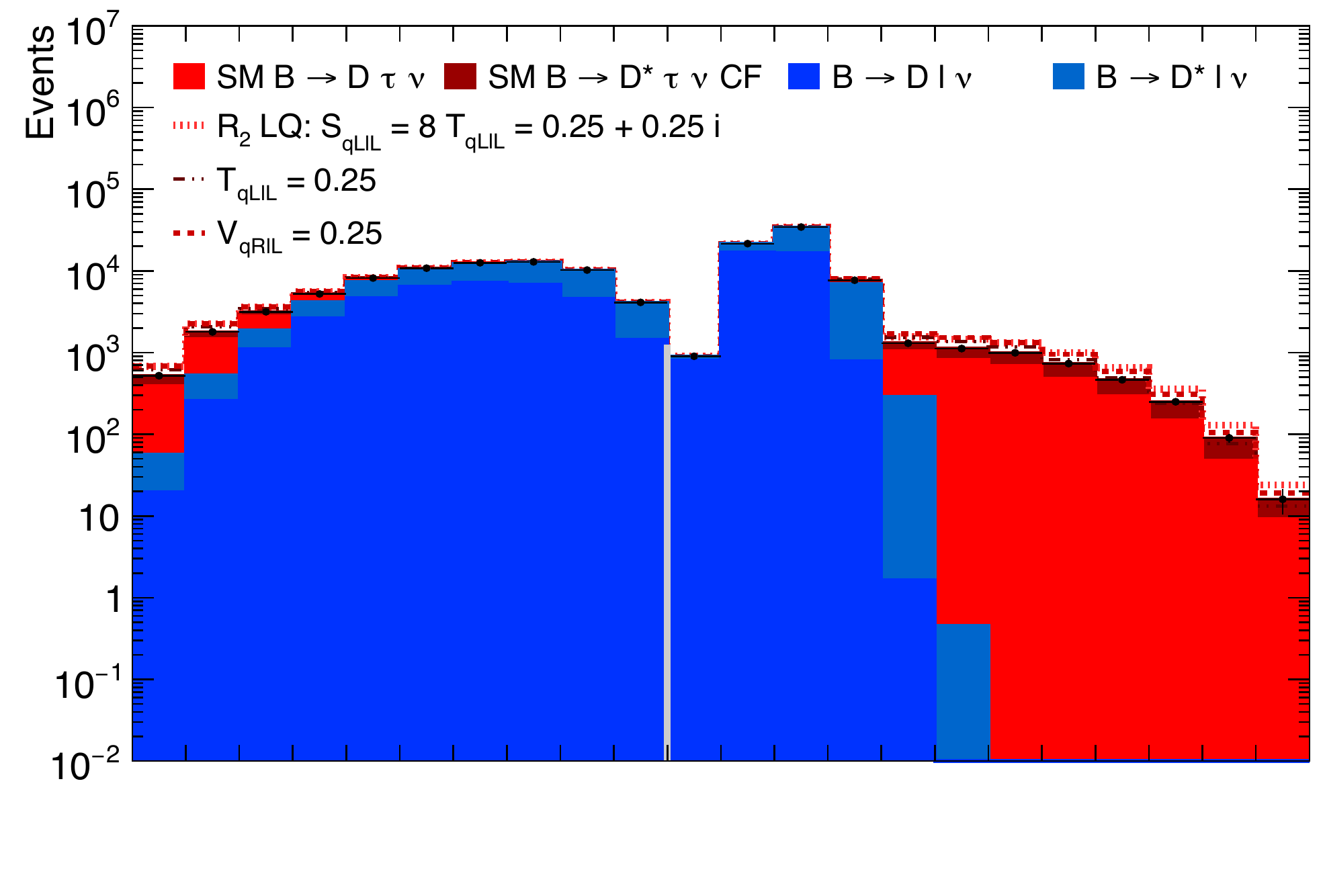}  \vspace{-4ex} \\ 
  \includegraphics[width=0.47\textwidth]{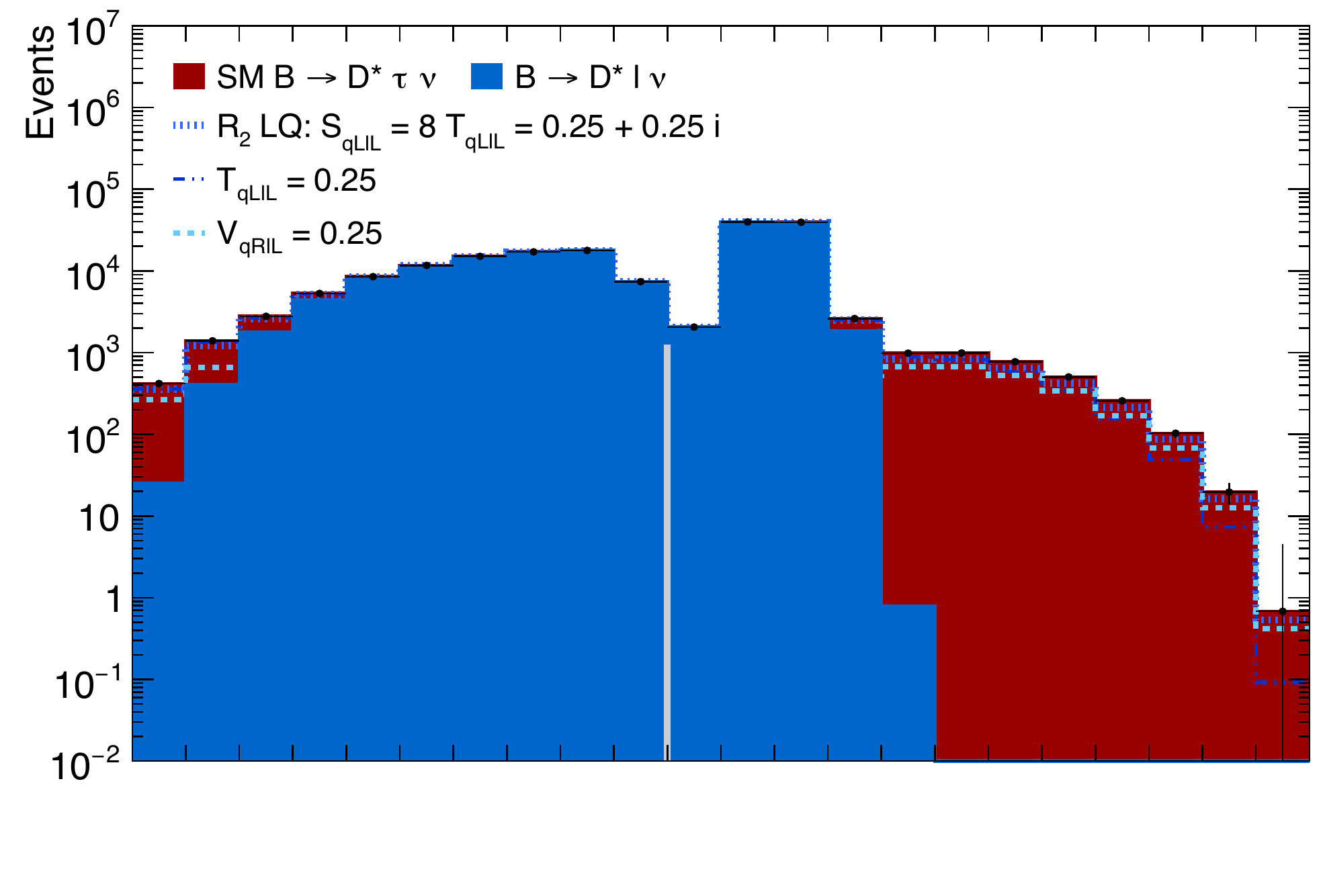}  \vspace{-5.8ex}  \\
  \includegraphics[width=0.47\textwidth]{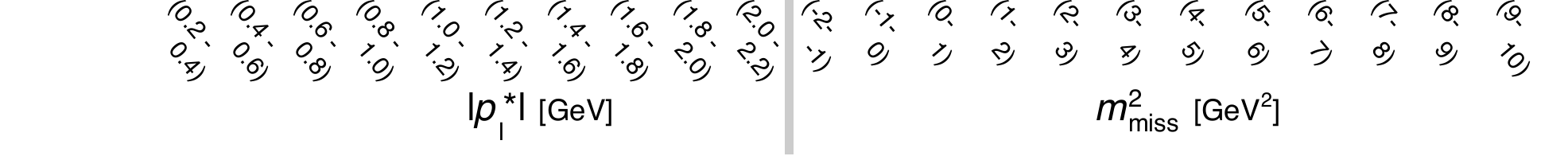}
\caption{ 
The $B \to D \, \tau \bar \nu$ (top) and $B \to D^* \tau \bar \nu$ (bottom) 
distributions in $|\vec p^*_{\ell}|$ and $m^2_{\text{miss}}$ in the Asimov data set. 
The number of events correspond to an estimated number of reconstructed events at Belle~II with $5 \, \text{ab}^{-1}$. 
}
\label{fig:NP_D_Ds}
\end{figure}

Figure~\ref{fig:NP_D_Ds} shows the projections of the constructed Asimov data set, as well as the distributions expected for the three NP models. 
The latter have the same couplings as those shown in Fig.~\ref{fig:NP_D_Ds_ratio}. 

\subsection{$R(D^{(*)})$ biases from new physics truth}
\label{subsec:NP_bias}

Many NP analyses and global fits to the $R(\dds)$ measurements -- together with other potentially template-sensitive observables, including $q^2$ spectra -- 
have been carried out by a range of phenomenological studies 
(see, e.g., Refs.~\cite{Dorsner:2016wpm, Freytsis:2015qca, Aebischer:2018iyb, Asadi:2019xrc, Bardhan:2019ljo, Bhattacharya:2018kig, Murgui:2019czp, Azatov:2018knx, Buttazzo:2017ixm, Blanke:2019qrx, Altmannshofer:2017poe, Becirevic:2018afm, Angelescu:2018tyl}).
As mentioned above, the standard practice has been to fit NP predictions to the world-average values of $R(\dds)$ (and other data) 
to determine confidence levels for allowed and excluded NP couplings.
However, because the $R(\dds)$ measurements use SM-based templates, 
and because the presence of NP operators can strongly alter acceptances and kinematic distributions, such analyses
can lead to incorrect best-fit values or exclusions of NP Wilson coefficients. 

To illustrate such a bias, we fit SM MC templates to NP Asimov data sets, that are generated with \hammer for three different NP `truth' 
benchmark points: the 2HDM Type~II with $S_{qRlL}  = -2$, corresponding to $\tan \beta / m_{H^+} \simeq 0.5$\,GeV$^{-1}$;
the same with $S_{qRlL}  = 0.75i$;
and the $R_2$ leptoquark model with $S_{qLlL} = 8 \, T_{qLlL} = 0.25 + 0.25\, i$.
(These models and couplings are for illustration; 
our goal here is only to demonstrate the type of biases that may plausibly be presumed to occur.) 
We replicate the fit of all existing measurements, allowing the normalizations of the $D$ and $D^*$ modes (and the light leptonic final states) to float independently, 
without imposing e.g. their predicted SM relationship. This fit leads to a best-fit ellipse in the $R(D)$\,--\,$R(D^*)$ plane. 

\begin{figure*}[ht!]
\centering
  \includegraphics[trim=10 0 5 0, clip, width=0.32\textwidth]{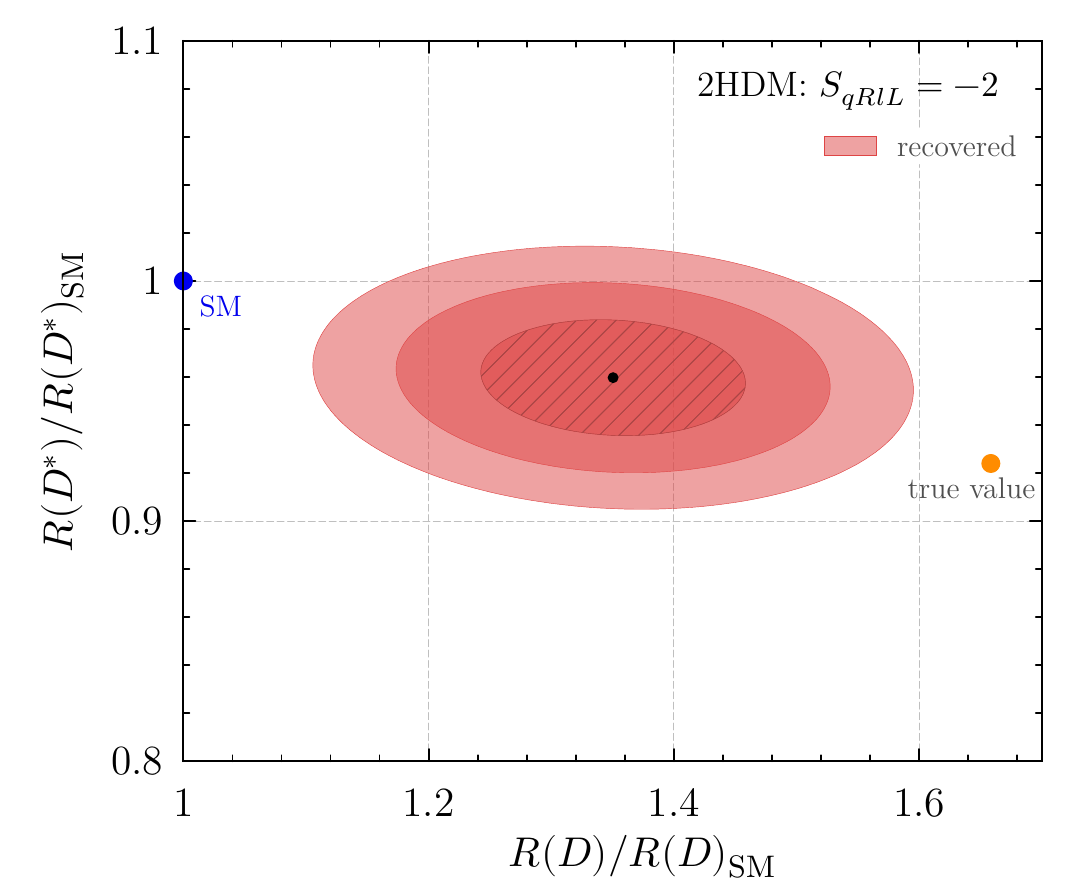} \hfil
  \includegraphics[trim=10 0 5 0, clip,width=0.32\textwidth]{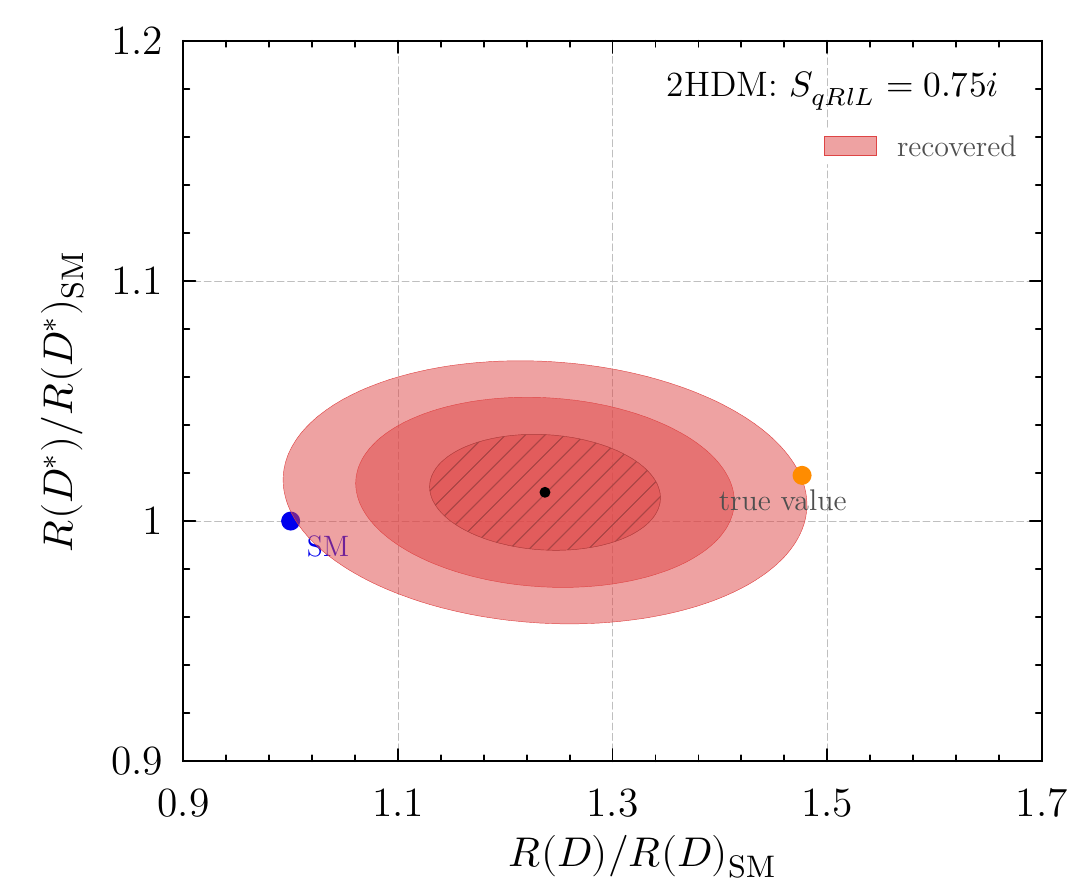} \hfil
  \includegraphics[trim=10 0 5 0, clip,width=0.32\textwidth]{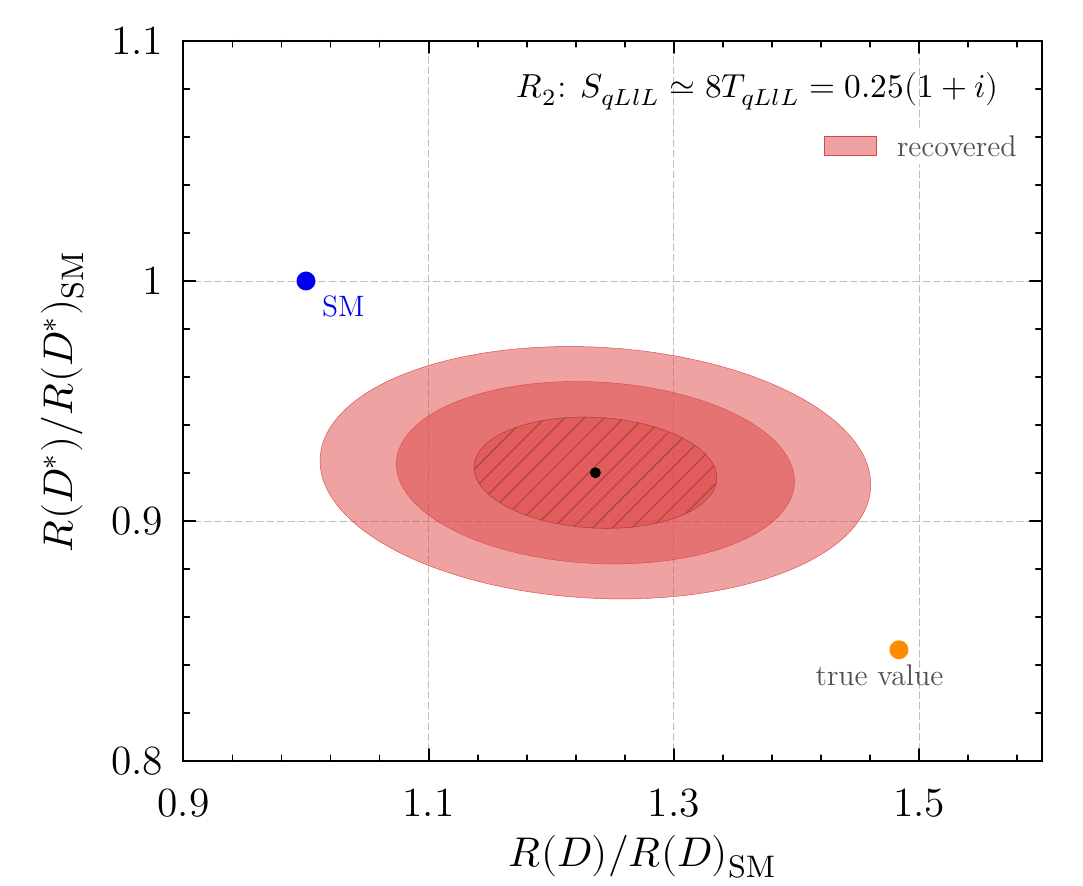} \\[4pt]
  \includegraphics[trim=10 0 5 0, clip,width=0.32\textwidth]{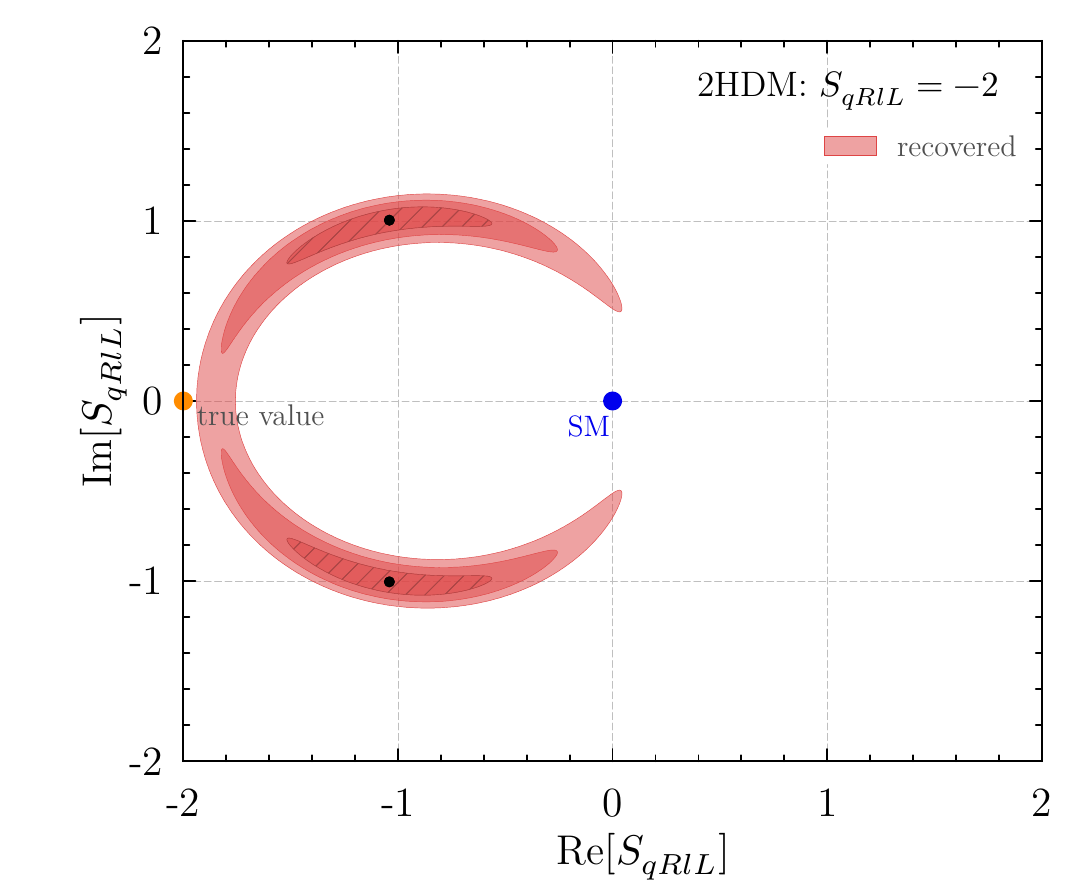} \hfil
  \includegraphics[trim=10 0 5 0, clip,width=0.32\textwidth]{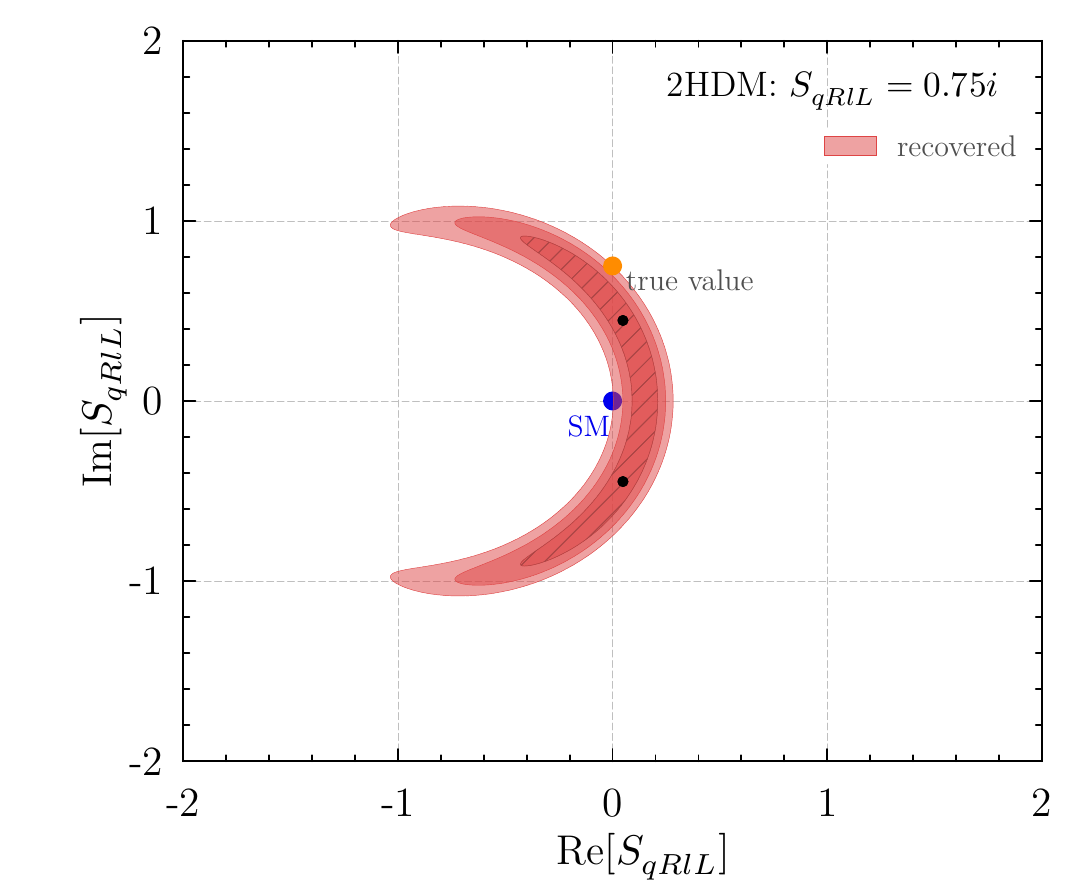} \hfil
  \includegraphics[trim=10 0 5 0, clip,width=0.32\textwidth]{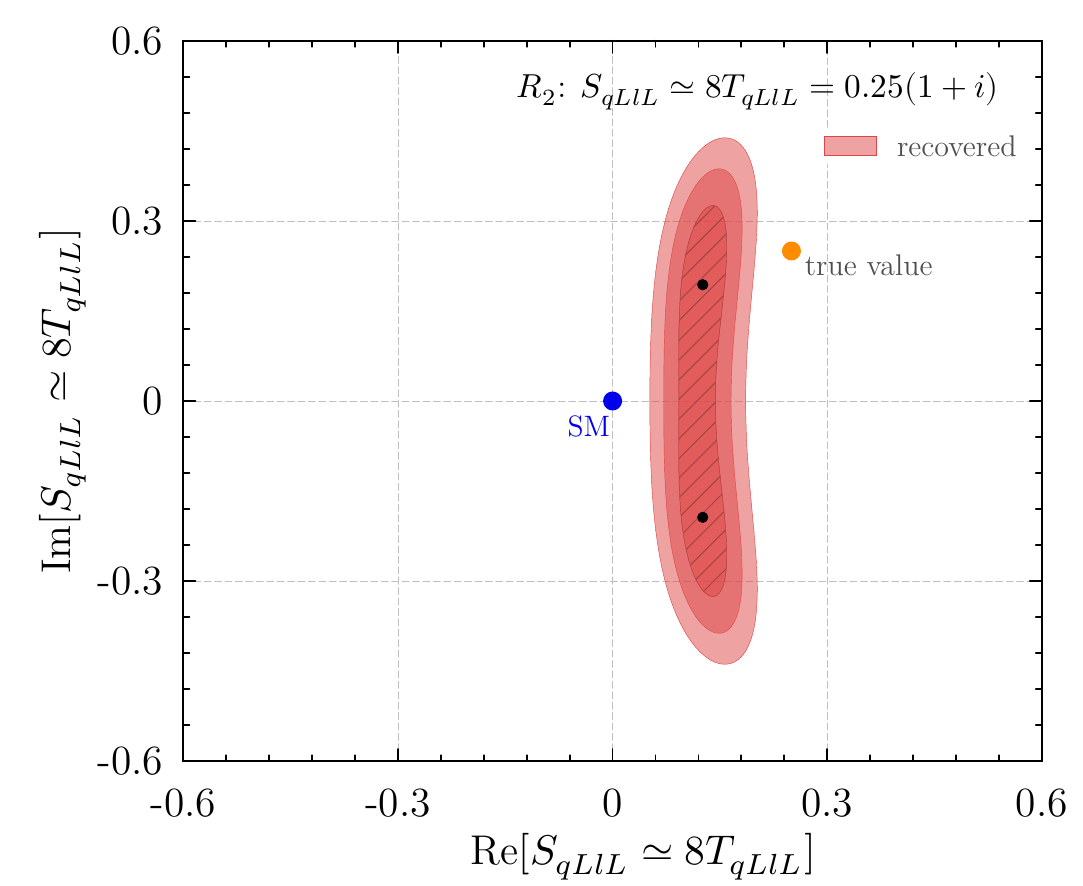} 
\caption{ 
 \emph{Top:} Illustrations of biases from fitting an SM template to three NP `truth' benchmark models: 
 the 2HDM type~II with $S_{qRlL}  = -2$ (left), $S_{qRlL}  = 0.75i$ (middle), and the $R_2$ leptoquark model with $S_{qLlL} = 8 \, T_{qLlL} = 0.25 + 0.25 i$ (right). 
 The orange dot corresponds to the predicted `true value' of $R(\dds)$ for the NP model, 
 to be compared to the recovered 68\%, 95\% and 99\% CLs of the SM fit to the NP Asimov data sets (with uncertainties estimated to correspond to $\sim5$\,ab$^{-1}$) in shades of red. 
  \emph{Bottom:} The best fit regions for the 2HDM and $R_2$ model Wilson coefficients obtained from fitting $R(\dds)$ NP predictions 
 to the recovered $R(\dds)$ CLs for each NP model. 
 The shades of red denote CLs as in the top row. The best fit (true value) Wilson coefficients are shown by black (orange) dots.
 }
\label{fig:NP_bias}
\end{figure*}

In Fig.~\ref{fig:NP_bias} we show the recovered values, $R(\dds)_{\text{rec}}$, obtained from this procedure, 
and compare them to the actual predictions of the given NP truth benchmark point, $R(\dds)_{\text{th}}$.
For ease of comparison, we normalize the $R(\dds)$ values against the SM predictions for $R(\dds)_{\text{SM}}$.
The resulting recovered best fit ratios, defining $\hat{R}(\dds) = R(\dds)/R(\dds)_{\mathrm{SM}}$
\begin{align}
	\text{2HDM ($-2$): ~} & \hat{R}(D)_{\text{rec}} = 1.35(7)\,, && \hat{R}(D)_{\text{th}} = 1.66 \nonumber\\
						& \hat{R}(D^*)_{\text{rec}} = 0.96(2)\,, && \hat{R}(D^*)_{\text{th}} = 0.92 \nonumber\\\
	\text{2HDM ($0.75i$):~} 	& \hat{R}(D)_{\text{rec}} = 1.24(7)\,, && \hat{R}(D)_{\text{th}} = 1.48 \nonumber\\
						& \hat{R}(D^*)_{\text{rec}} = 1.01(2)\,, && \hat{R}(D^*)_{\text{th}} = 1.02 \nonumber\\\					
	\text{$R_2$:~} 		& \hat{R}(D)_{\text{rec}} = 1.24(7)\,, && \hat{R}(D)_{\text{th}} = 1.48 \nonumber\\
						& \hat{R}(D^*)_{\text{rec}} = 0.92(2)\,, && \hat{R}(D^*)_{\text{th}} = 0.85\,. \nonumber						
\end{align}
For two NP models, the recovered ratios from fitting the Asimov data set exclude the truth $R(\dds)_{\text{th}}$ values at $\gtrsim 4\sigma$, and the other at $3\sigma$.
The recovered ratios show deviations from the SM comparable in size (but in some cases a different direction) to the current world average $R(\dds)$, 
and much smaller than the deviations expected from the truth $R(\dds)_{\text{th}}$ values. 
This illustrates the sizable bias in the measured $R(\dds)$ values that may be presumed to ensue from carrying out fits with an SM template, if NP actually contributes to the measurements. 
We emphasize that the degree to which a particular NP model is actually affected by this type of bias -- including the size and direction of the bias -- 
may be sensitive to the details of the experimental framework 
and is therefore a question that can only be answered within each experimental analysis.

We also show in Fig.~\ref{fig:NP_bias} the equivalent bias arising from a na\"\i ve fit of the $R(\dds)$ NP prediction that attempts to recover the complex Wilson coefficient.
This is done by parametrizing $R(\dds)_{\text{th}} = R(\dds)[c_{XY}]$, and fitting this expression to the recovered $R(\dds)_{\text{rec}}$ values.
Explicitly, one calculates CLs in the Wilson coefficient space via the two degree of freedom chi-square $\chi^2 = \bm{v}^T \sigma_{R(\dds)}^{-1} \bm{v}$, with
$ \bm{v} = \big(R(D)_{\text{th}} - R(D)_{\text{rec}}\,, R(D^*)_{\text{th}} - R(D^*)_{\text{rec}}\big)$.
The resulting best fit Wilson coefficient regions similarly exclude the truth values.

Thus, the allowed or excluded regions of NP couplings determined from fits to the $R(\dds)$ measurements must be treated with caution, 
as these fits do not include effects of the NP distributions in the MC templates.
Similarly, results of global fits should be interpreted carefully when assessing the level of compatibility with specific NP scenarios. 

\subsection{New physics Wilson coefficient fits}

Instead of considering observables like $R(\dds)$, for phenomenological studies to be able to properly make interpretations and test NP models, 
experiments should provide direct constraints on NP Wilson coefficients themselves.
For example, this could be done with simplified likelihood ratios that profile out all irrelevant nuisance parameters 
from, e.g., systematic uncertainties or information from sidebands or control channels, or by other means.

\begin{figure}[tbh]
\centering
  \includegraphics[width=0.45\textwidth]{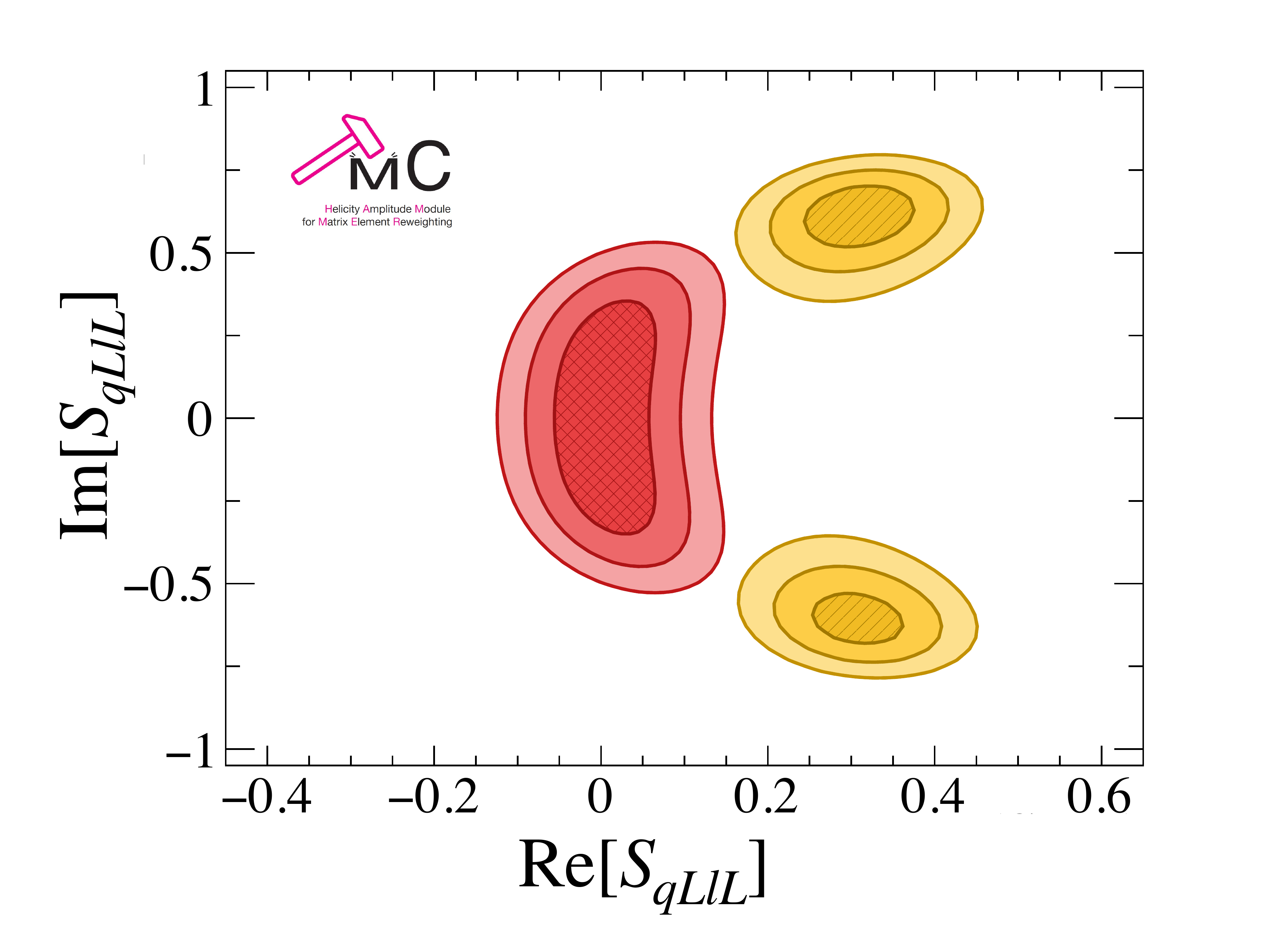} \\
  \includegraphics[width=0.45\textwidth]{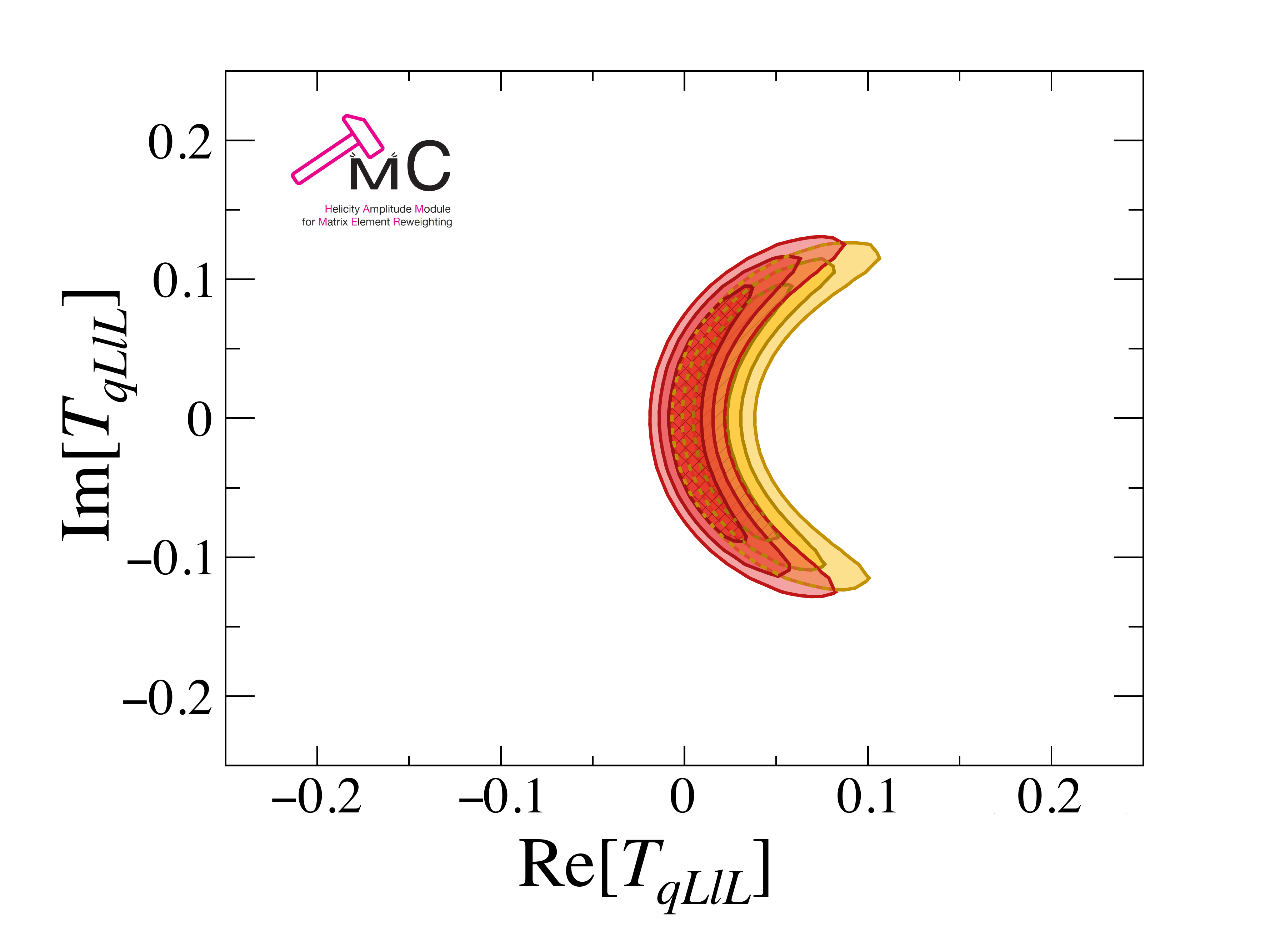} \\
  \includegraphics[width=0.45\textwidth]{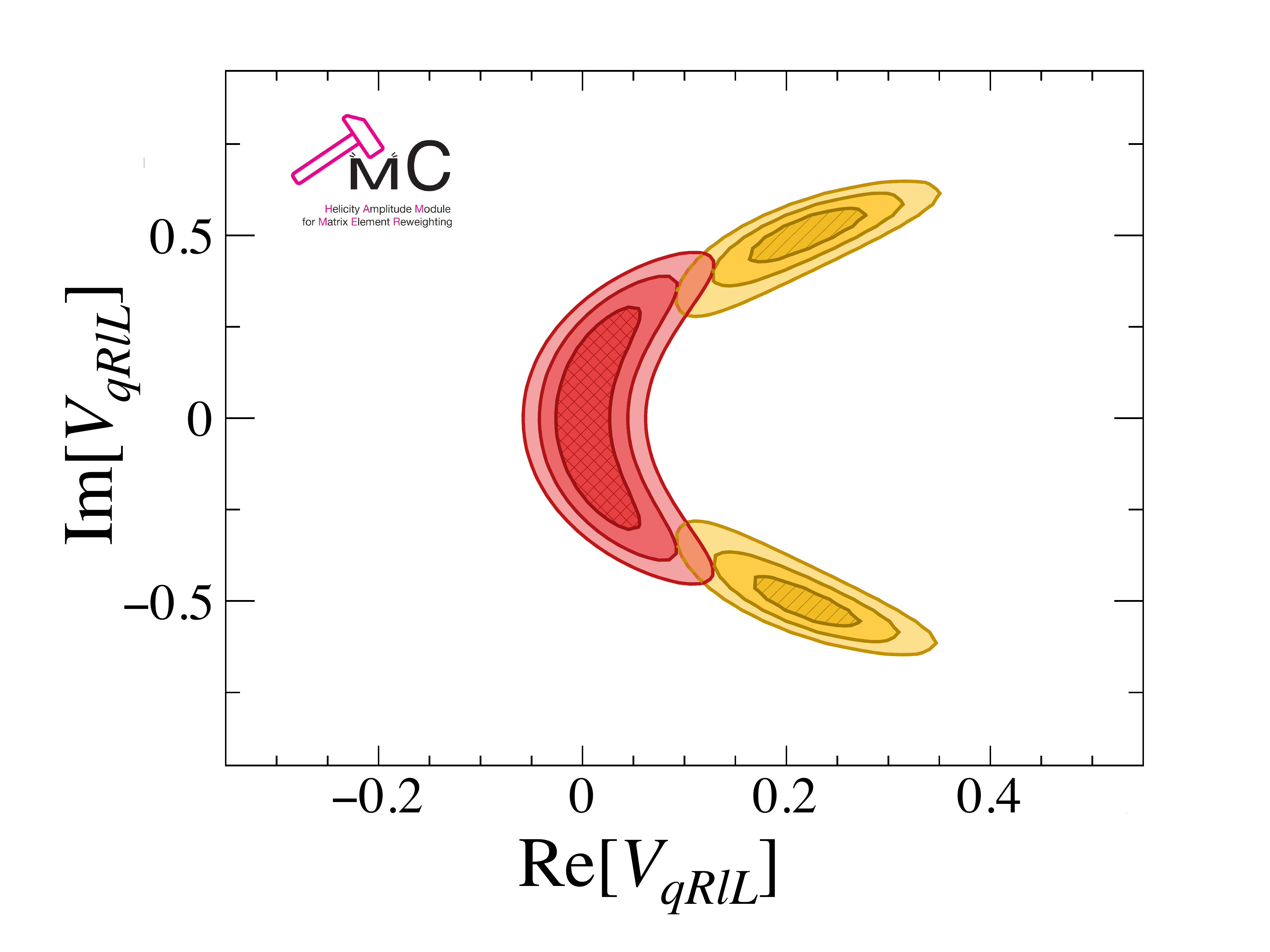}   
\caption{The 68\%, 95\% and 99\% CL allowed regions of the three models under consideration, from fitting the SM (red) and 2HDM type~II (yellow and with $S_{qRlL}  = -2$) Asimov data sets. (Top) $R_2$ leptoquark model with $S_{qLlL} = 8 T_{qLlL}$; (middle) NP in the form of a left-handed tensor coupling; (bottom) NP in the form of a right-handed vector coupling. 
 }
\label{fig:NP_Scan}
\end{figure}

As an example, we now use \hammer to perform such a fit for the real and imaginary parts of the NP Wilson coefficients, 
using the set of three NP models in Sec.~\ref{sec:rwgt} as templates. 
These are fit to the same two truth benchmark scenarios as in Fig.~\ref{fig:NP_Scan}: 
a truth SM Asimov data set;
and a truth Asimov data set reweighted to the 2HDM Type II with $S_{qRlL}  = -2$. 

Figure~\ref{fig:NP_Scan} shows in shades of red the 68\%, 95\% and 99\% confidence levels (CLs) of the three NP model scans of SM Asimov data sets. 
For the SM truth benchmark, the corresponding best fit points are always at zero NP couplings. 
The derived CLs then correspond to the expected median exclusion of the fitted NP coupling under the assumption the SM is true. 

We further show in shades of yellow the same fit CLs for the 2HDM truth benchmark Asimov data set.
These latter fits illustrate a scenario in which NP is present, but is analyzed with an incomplete or incorrect set of NP Wilson coefficients. 
Depending on the set of coefficients, we see from the $\Delta \chi^2$ of the best fit points that the new physics might be obfuscated or wrongly identified. 
This underlines the importance for LHCb and Belle~II to eventually carry out an analysis in the full multi-dimensional space of Wilson coefficients, spanned by the operators listed in Table~\ref{tab:NPc}. 

\section{The \hammer library}
\label{sec:code}

In this section we present core interface features and calculational strategies of the \hammer library. 
Details of the code structure, implementation, and use, can be found in the \hammer manual~\cite{hammer_manual}; here we provide only an overview.  

\subsection{Reweighting}

We consider an MC event sample, comprising a set of events indexed by $I$, with weights $w_I$ and truth-level kinematics $\{q\}_I$.
Reweighting this sample from an `old' to a `new'  theory requires the truth-level computation of the ratio of the differential rates
\begin{equation}
	\label{eqn:renorm}
	r_I = \frac{ d \Gamma^{\text{new}}_I/d \PSp }{d \Gamma^{\text{old}}_I/d \PSp}\,,
\end{equation}
applied event-by-event via the mapping $w_I \mapsto r_I w_I$. The `old' or `input' or `denominator' theory is typically the SM plus (where relevant) 
a hadronic model --- that is, a form factor (FF) parametrization. (It may also be composed of pure phase space (PS) elements, see App.~\ref{app:spec}.)
The `new' or `output' or `numerator' theory may involve NP beyond the Standard Model, or a different hadronic model, or both.

Historically, the primary focus of the library is reweighting of $b \to c\ell\nu$ semileptonic processes, often in multistep cascades such as $B \to D^{(*,**)}(\to DY)\, \tau(\to X\nu) \bar\nu$. 
However, the library's computational structure is designed to be generalized beyond these processes, 
and we therefore frame the following discussion in general terms, before returning to the specific case of semileptonic decays.

\subsection{New Physics generalizations}
\label{sec:NPgen}

The \hammer library is designed for the reweighting of processes via theories of the form
\begin{equation}
	\mathcal{L} = \sum_\alpha c_\alpha\, \mathcal{O}_\alpha\,.
\end{equation}
where $\mathcal{O}_\alpha$  are a basis of operators, and $c_\alpha$, are SM or NP Wilson coefficients 
(defined at a fixed physical scale; mixing of the Wilson coefficients under RG evolution, if relevant, must be accounted for externally to the library).
We specify in Table~\ref{tab:NPc} the conventions used for various $b \to c\ell\nu$ four-Fermi operators and other processes included in the library.

The corresponding process amplitudes may be expressed as linear combinations $c_\alpha \mathcal{A}_\alpha$. 
They may also be further expressed as a linear sum with respect to a basis of form factors, $F_i$, that encode the physics of hadronic transitions (if any).\footnote{In all $b \to c$ processes currently handled by \hammer\ (see Table~\ref{tab:knownampls} for a list) 
the form factors are functions of $q^2 = \big(p_{H_b} -p_{H_c}\big)^2$,
or equivalently of the dimensionless kinematic variable,
\begin{equation}
\label{eqn:wdef}
	w = v\cdot v' = \frac{m_{H_b} ^2  + m_{H_c}^2 - q^2}{2m_{H_b}  m_{H_c}}\,,
\end{equation}
with four velocities $v = p_{H_b}  / m_{H_b} $ and $v' = p_{H_c} / m_{H_c}$.
For decays with multi-hadron final states, such as the $\tau \to n \pi$, $n\ge 3$, the form factors are also dependent on multiple invariant masses of the final state hadrons.
Thus, $b \to c\tau\nu$ decays followed by with subsequent hadronic $\tau$ decays involve at least two separate sets of hadronic functions at the amplitude level.}
In general, then, an amplitude may be written in the form
\begin{equation}
	\mathcal{M}^{\{s\}}\big(\{q\}\big) = \sum_{\alpha, i} c_\alpha \, F_i\big(\{q\}\big) \, \mathcal{A}^{\{s\}}_{\alpha i}\big(\{q\}\big)\,,
\end{equation}
in which $\{s\}$ are a set of external quantum numbers and $\{q\}$ the set of four-momenta.\footnote{
The momenta of an event passed to the library must all be specified in the same frame. The choice of frame is arbitrary.} 
The object $\mathcal{A}_{\alpha i}$ is an NP- and FF-generalized \emph{amplitude tensor}.  
In the case of cascades, relevant for $B \to D^{(*,**)}(\to DY)\, \tau(\to X\nu) \bar\nu$ decays, 
the amplitude tensor may itself be the product of several subamplitudes, summed over several sets of internal quantum numbers.
The corresponding polarized differential rate
\begin{align}
	\label{eqn:WgtTen}
	\frac{d\Gamma^{\{s\}}}{d\PSp} & = \!\! \sum_{\alpha, i, \beta, j} \!\! c_\alpha c_\beta^\dagger \, F_i F_j^\dagger\!\big(\{q\}\big) \, \mathcal{A}^{\{s\}}_{\alpha i}\mathcal{A}^{\dagger\{s\}}_{\beta j}\!\big(\{q\}\big) \,, \notag \\
	& = \!\! \sum_{\alpha, i, \beta, j} \!\! c_\alpha c_\beta^\dagger \, F_i F_j^\dagger\!\big(\{q\}\big) \,\mathcal{W}_{\alpha i \beta j}\,,
\end{align}
in which the phase space differential form $d\PSp$ includes on-shell $\delta$-functions and geometric or combinatoric factors, as appropriate.

The outer product of the amplitude tensor, defined as $\mathcal{W} \equiv \mathcal{A} \mathcal{A}^\dagger$, is a \emph{weight tensor}. 
The object $\sum_{ij} F_i F_j^\dagger \mathcal{W}_{\alpha i \beta j}$ in Eq.~\eqref{eqn:WgtTen} is independent of the Wilson coefficients:
Once this object is computed for a specific $\{q\}$ -- an event -- it can be contracted with any choice of NP to generate an event weight. 
Similarly, on a patch of phase space $\Omega$ --- e.g., the acceptance of a detector or a bin of a histogram --- the marginal rate can now be written as
\begin{equation}
	\label{eqn:Gsom}
	\Gamma^{\{s\}}_{\Omega} = \sum_{\alpha, \beta} c_\alpha c_\beta^\dagger \int_\Omega d\PSp \, \sum_{ij} F_i F_j^\dagger\big(\{q\}\big)  \mathcal{W}^{\{s\}}_{\alpha i \beta_j} \big(\{q\}\big)\,.
\end{equation}
The Wilson coefficients factor out of the phase space integral, so that the integral itself generates a NP-generalized tensor. 
After it is computed once, it can be contracted with any choice of NP Wilson coefficients, $c_\alpha$, thereafter.

The core of \hammer's computational philosophy is based on the observation that this contraction is computationally much more efficient than the initial computation (and integration).
Hence efficient reweighting is achieved by
\begin{itemize}
	\item Computing NP (and/or FF, see below) generalized objects, and storing them;
	\item Contracting them thereafter for any given NP (and/or FF) choice to quickly generate a desired NP (and/or FF) weight.
\end{itemize}

\begin{table}[tb]
\begin{center}
\renewcommand{\arraystretch}{1.05}
\newcolumntype{D}{ >{\centering\arraybackslash} m{0.31\linewidth} <{}}
\newcolumntype{E}{ >{\centering\arraybackslash\ttfamily} m{0.55\linewidth} <{}}
\newcommand{\citerm}[1]{{\rmfamily{\cite{#1}}}}
\begin{tabular}{D|E}
\hline\hline
Process &  \textrm{Form factor parametrizations}\\
\hline
$B \to D^{(*)} \ell \nu$  &  \textlst{ISGW2}$^*$\,\citerm{Scora:1995ty,Isgur:1988gb}, \textlst{BGL}$^{*}$\,\citerm{Grinstein:2017nlq, Boyd:1995sq, Boyd:1997kz}, \textlst{CLN}$^{*\ddagger}$\,\citerm{Caprini:1997mu}, \textlst{BLPR}$^{\ddagger}$\,\citerm{Bernlochner:2017jka}\\
$B \to (D^* \to D \pi) \ell \nu$  & \textlst{ISGW2}$^*$, \textlst{BGL}$^{*\ddagger}$, \textlst{CLN}$^{*\ddagger}$, \textlst{BLPR}$^{\ddagger}$\\
$B \to (D^* \to D\gamma) \ell \nu$  &  \textlst{ISGW2}$^*$, \textlst{BGL}$^{*\ddagger}$, \textlst{CLN}$^{*\ddagger}$, \textlst{BLPR}$^{\ddagger}$\\
\hline
$\tau \to \pi \nu$ & \textrm{---}\\
$\tau \to \ell \nu \nu$ &   \textrm{---}\\
$\tau \to 3\pi \nu$ &  \textlst{RCT}$^*$\,\citerm{Kuhn:1992nz,Shekhovtsova:2012ra,Nugent:2013hxa}\\
\hline
$ B \to D^*_0 \ell \nu$ &  \textlst{ISGW2}$^*$, \textlst{LLSW}$^*$\,\citerm{Leibovich:1997em, Leibovich:1997tu}, \textlst{BLR}$^{\ddagger}$\,\citerm{Bernlochner:2017jxt, Bernlochner:2016bci}\\
$ B \to D^*_1 \ell \nu$ &  \textlst{ISGW2}$^*$, \textlst{LLSW}$^*$, \textlst{BLR}$^{\ddagger}$ \\
$ B \to D_1 \ell \nu$ &  \textlst{ISGW2}$^*$, \textlst{LLSW}$^*$, \textlst{BLR}$^{\ddagger}$ \\
$ B \to D^*_2 \ell \nu$ &  \textlst{ISGW2}$^*$, \textlst{LLSW}$^*$, \textlst{BLR}$^{\ddagger}$ \\
\hline
$\Lambda_b \to \Lambda_c \ell \nu$ & \textlst{PCR}$^*$\,\citerm{Pervin:2005ve}, \textlst{BLRS}$^{\ddagger}$\,\citerm{Bernlochner:2018bfn, Bernlochner:2018kxh} \\
\hline
\multicolumn{2}{c}{Planned for next release}\\
\hline
$ B_{(c)} \to \ell \nu$ &  \textlst{MSbar}\\
$ B \to (\rho \to \pi\pi)\ell \nu$ & \textlst{BCL}$^*$\,\citerm{Bourrely:2008za}, \textlst{BSZ}\,\citerm{Straub:2015ica}\\
$ B \to (\omega \to \pi\pi\pi)\ell \nu$ &  \textlst{BCL}$^*$, \textlst{BSZ} \\
$ B_c \to (J\!/\!\psi \to \ell\ell)\ell \nu$ & \textlst{EFG}$^*$\,\citerm{Ebert:2003cn}, \textlst{BGL}$^{*\ddagger}$\,\cite{Cohen:2019zev}\\
\hline
$ \Lambda_b \to \Lambda_c^* \ell \nu$ & \textlst{PCR}$^*$\,, ... \\
\hline
$\tau \to 4\pi \nu$ &  \textlst{RCT}$^*$\\
$\tau \to (\rho \to \pi\pi)\nu$ &  \textrm{---}\\
\hline\hline
\end{tabular}
\caption{Presently implemented amplitudes in the \hammer library, and corresponding form factor parametrizations. SM-only parametrizations are indicated by a $*$ superscript. Form factor parametrizations that include linearized variations are denoted with a $\ddagger$ superscript. These are named in the library by adding a ``\textlst{Var}'' suffix, e.g. ``\textlst{BGLVar}''.}
\label{tab:knownampls}
\end{center}
\end{table}

\subsection{Form factor generalizations}

Similarly to the NP Wilson coefficients, it is often desirable to be able to vary the FF parameterizations themselves. 
This can be achieved directly within \texttt{Hammer} by adjusting the choice of FF parameter values for any given parametrization.
However, because the impacts of the form-factors depend on the kinematics of an event, they cannot be factored out of the phase-space integral in Eq.~\eqref{eqn:Gsom}.
Full reweighting to a different choice of form-factor parameters therefore requires full recalculation of each event weight on the phase space patch.

Instead, one might contemplate linearized variations with respect to the FF parameters, that commute with the phase space integration:
For instance, variations around a (best-fit) point along the error eigenbasis of a fit to the FF parameters; or FF parametrizations that are linearized with respect to a basis of parameters, 
such as the BGL parametrization~\cite{Grinstein:2017nlq, Boyd:1995sq, Boyd:1997kz} in $B \to D^{(*)} \ell \nu$. 
To this end, an FF parametrization with a parameter set $\{\mu\}$ can be linearized around a (best-fit) point, $\{\mu^0\}$ so that 
\begin{equation}
	\label{eqn:FFerr}
	F_{i}\big(\{q\}; \{\mu\}\big) = F_i\big(\{q\}, \{\mu^0\}\big) + \sum_{a} F'_{i,a}\big(\{q\}, \{\mu^0\}\big)\,e_a\,,
\end{equation}
where `$a$' is one or more \emph{variational indices} and $e_{a}$ is the variation. 
In the language of the error eigenbasis case, $F'_{i,a}$ is the perturbation of $F_i$ in the $a$th principal component $e_{a}$ of the parametric fit covariance matrix.

Defining $\xi_a \equiv (1, e_a)$ and $\Phi_{i,a+1} \equiv (F_{i}, F'_{i,a})$, so that Eq.~\eqref{eqn:FFerr} becomes
\begin{equation}
	\sum_{a} \xi_a \Phi_{i,a} = F_{i} + \sum_{a'} F'_{i,a'}\,e_{a'}\,,
\end{equation}
then the differential rate
\begin{multline}
	\label{eqn:GWgtTen}
	\frac{d\Gamma^{\{s\}}}{d\PSp} = \!\! \sum_{\alpha, a, \beta, b} \!\! c_\alpha c_\beta^\dagger \xi_a \xi^\dagger_b \mathcal{U}^{\{s\}}_{\alpha a \beta b}\,,\\
	\mathcal{U}^{\{s\}}_{\alpha a \beta b} \equiv \sum_{ij} \Phi_{i,a} \Phi_{j,b}^\dagger\big(\{q\}\big)  \mathcal{W}^{\{s\}}_{\alpha i \beta j} \big(\{q\}\big)\,,
\end{multline}
with $\mathcal{U}$ an NP- and FF-generalized weight tensor. The $\xi_a$ are independent of $\{q\}$ and factor out of any phase space integral just as the Wilson coefficients do. That is, an integral on any phase space patch,
\begin{equation}
	\Gamma^{\{s\}}_{\Omega} = \sum_{\alpha, \beta, a, b} c_\alpha c_\beta^\dagger \xi_a \xi^\dagger_b \int_\Omega d\PSp \,\, \mathcal{U}^{\{s\}}_{\alpha a \beta b}\,. \end{equation}
One may thus tensorialize the amplitude with respect to Wilson coefficients and/or FF linearized variations, to be contracted later with with NP or FF variation choices 
(the latter within the regime of validity of the FF linearization).
Hereafter, the $\xi_a$ are referred to as `FF uncertainties' or `FF eigenvectors' following the nominal fit covariance matrix example. 

\subsection{Rates}

In certain cases, it is also useful to compute and fold in an overall ratio of rates $\Gamma^{\text{old}}/\Gamma^{\text{new}}$, 
or the rates themselves, $\Gamma^{\text{new}, \text{old}}$, may be required. 
For example, if the MC sample has been initially generated with a fixed overall branching ratio, $\BR_{\text{new}}$, 
one might wish to enforce this constraint via an additional multiplicative factor $\BR_{\text{old}}/\BR_{\text{new}}$.

The different components computed by \hammer are then: 
\begin{enumerate}[(i)]
\item The NP- and/or FF-generalized tensor for $(d \Gamma^{\text{new}}_I/d \PSp) / (d \Gamma^{\text{old}}_I/d \PSp)$, 
via Eq.~\eqref{eqn:GWgtTen}, noting that the denominator carries no free NP or FF variational index. (The ratio $r_I$ is then itself generally at least a rank-2 tensor.); 
\item The NP- and/or FF-generalized \emph{rate tensors} $\Gamma^{\text{old, new}}$, which need be computed only once for an entire sample.  
(These rates require integration over the phase space, which is achieved by a dedicated multidimensional Gaussian quadrature integrator.)
\end{enumerate}

\subsection{Primary code functionalities}

The calculational core of \hammer computes the NP or FF generalized tensors event-by-event for any process known to the library (see Table~\ref{tab:knownampls} for a list), 
and as specified by initialization choices (more detail is provided in Sec.~\ref{sec:code_flow}) and specified form factor parametrizations.
This core is supplemented by a wide array of functionalities to permit manipulation the resulting NP- and FF-generalized weight tensors as needed. 
This may include binning --- equivalent to integrating on a phase space patch --- the weight tensors into a histogram of any desired reconstructed observables, 
and/or it may include folding of detector simulation smearings, etc.
Such histograms have NP- and FF-generalized tensors as bin entries, and we therefore call them \emph{generalized} or \emph{tensor} histograms.
Once such NP- and FF-generalized tensor objects are computed and stored, contraction with NP or FF eigenvector choices 
permits the library to efficiently generate actual event weights or histogram bin weights for any theory of interest.

The architecture of \hammer is designed around several primary functionalities:
\begin{enumerate}
	\item Provide an interface to determine which processes are to be reweighed, and which (possibly multiple) schemes for form factor parametrizations are to be used. 
	This includes handling for (sub)processes that were generated as pure phase space, and the ability to change the values of the form factor parameters.
	\item Parse events into cascades of amplitudes known to the library, and compute their corresponding NP- and/or FF-generalized amplitude or weight tensor, 
	as well as the respective rate tensors, as needed.
	\item Provide an interface to generate histograms (of arbitrary dimension), and bin the event weight tensors --- i.e., $r_I w_I$, as in Eq.~\eqref{eqn:renorm} --- 
	into these histograms, as instructed. 
	This includes functionality for weight-squared statistical errors, functionality for generation of \texttt{ROOT} histograms, 
	as well as extensive internal architecture for efficient memory usage.
	\item Efficiently contract generalized weight tensors or bin entries against specific FF variational or NP choices, to generate an event or bin weight. 
	This includes extensive internal architecture to balance speed versus memory requirements.
	\item Provide interface to save and reload amplitude or weight tensors or generalized histograms, 
	to permit quick reprocessing into weights from precomputed or `initialized' tensor objects. 
\end{enumerate}
Examples of the implementation of these functionalities are shown in many examples provided with the source code.

\subsection{Code flow}
\label{sec:code_flow}

A \hammer program may have two different types of structure: An \emph{initialization} program, so called as it runs on MC as input, and may generate \hammer format files; 
or an \emph{analysis} program, which may reprocess histograms or event weights that have already been saved in an initialization run. 
Pertinent details of the elements of the application programming interface mentioned below are provided in Appendix~\ref{sec:API}, with more details in the \hammer manual. 

An initialization program has the generic flow:
\begin{enumerate}
	\item Create a \hammer object.
	\item Declare included or forbidden processes, via \textlst{includeDecay} and \textlst{forbidDecay}.
	\item Declare form factor schemes, via \textlst{addFFScheme} and \textlst{setFFInputScheme}.
	\item (Optional) Add histograms, via \textlst{addHistogram}.
	\item (Optional) Declare the MC units, via \textlst{setUnits}. 
	\item Initialize the \hammer class members with \textlst{initRun}.
	\item (Optional) Change FF default parameter settings with \textlst{setOptions}, or (if not SM) declare the Wilson coefficients for the input MC via \textlst{setWilsonCoefficients}.
	\item (Optional) Fix Wilson coefficient (Wilson coefficient and/or FF uncertainty) choice to special choices in weight calculations (histogram binnings), via
	\textlst{specializeWCInWeights} (\textlst{specializeWCInHistogram} and/or \textlst{specializeFFInHistogram}).
	\item Each event may contain multiple processes, e.g., a signal and tag $B$ decay. Looping over the events:
		\begin{enumerate}
			\item Initialize event with \textlst{initEvent}. For each process in the event:
			\begin{enumerate}
				\item Create a \hammer \textlst{Process} object.
				\item Add particles and decay vertices to create a process tree, via \textlst{addParticle} and \textlst{addVertex}.
				\item Decide whether to include or exclude processes from an event via \textlst{addProcess} and/or \textlst{removeProcess}.
			\end{enumerate}
			\item Compute or obtain event observables -- specific particles can be extracted with \textlst{getParticlesByVertex} or other programmatic means -- 
			and specify the corresponding histogram bins to be filled via \textlst{fillEventHistogram}.
			\item Initialize and compute the process amplitudes and weight tensors for included processes in the event, and fill histograms with event tensor weights -- 
			the direct product of include process tensor weights -- via \textlst{processEvent}.
			\item  (Optional) Save the weight tensors for each event, with \textlst{saveEventWeights} to a buffer.
		\end{enumerate}
	\item (Optional) Generate histograms with \textlst{getHistogram(s)} and/or save them with \textlst{saveHistograms}. NP choices are implemented with \textlst{setWilsonCoefficients}, FF variations are set with \textlst{setFFEigenvectors}.
	\item  (Optional) Save the rate tensors, with \textlst{saveRates} to a buffer.
	\item (Optional) Save an autogenerated \texttt{bibTeX} list of references used in the run with \textlst{saveReferences}.
\end{enumerate}

By contrast, an \emph{analysis} program (from a previously initialized sample, stored in a buffer) has the generic flow:
\begin{enumerate}
	\item Create a \hammer object and specify the input file.
	\item Load or merge the run header --- include or forbid specifications, FF schemes, or histograms --- with \textlst{loadRunHeader} (after \textlst{initRun}). One may further declare additional histograms to be compiled (from saved event weight data) via \textlst{addHistogram}.
	\item (Optional) Load or merge saved histograms with \textlst{loadHistograms}, and/or generate desired histograms with \textlst{getHistogram(s)}. NP choices are implemented with \textlst{setWilsonCoefficients}.
	\item (Optional) Looping over the events:
		\begin{enumerate}
			\item Initialize event with \textlst{initEvent}.
			\item If desired, remove processes from an event with \textlst{removeProcess}.
			\item Reload event weights with \textlst{loadEventWeights}.
			\item Specify histograms to be filled via \textlst{fillEventHistogram}.
			\item Fill histograms with event weights via \textlst{processEvent}.
		\end{enumerate}
\end{enumerate}

\section{Conclusions}
\label{sec:summary}

Precision measurements of $b\to c \tau \bar \nu$ decays require large Monte Carlo samples, which incorporate detailed simulations of detector responses and physics backgrounds.
The limited statistics due to the computational cost of these simulations are often a leading systematic uncertainty in the measurements, 
and it is prohibitively expensive to generate fully simulated MC samples for arbitrary NP models or descriptions of hadronic matrix elements. 

In this paper we described the \hammer library, and illustrated its utility. \hammer\ allows the fast and efficient reweighting of existing 
SM (or phase-space based) MC samples to arbitrary NP models. 
In addition, \hammer\ can be used to change form factor parametrizations and/or incorporate uncertainties from form factors into experimental measurements.
\hammer\ provides a computationally fast way for binned fits to generate predictions, 
and we implement a demonstrative forward-folding fit to constrain NP Wilson coefficients using this feature. 
Such a fit should be carried out by experimental collaborations in future measurements to provide reliable constraints on NP contributions in semileptonic $b\to c \tau \bar \nu$ decays. 
The results will allow people outside the collaborations to make correct interpretations of the data, which has not been possible to date without potentially sizeable biases. 
To demonstrate this latter point, we carried out toy NP analyses using SM fits to NP Asimov data sets, and showed that sizeable biases can indeed occur. 
\hammer\ is open source software and we are looking forward to the experimental results and interpretations it will enable.

\begin{acknowledgements} \phantom{a}

\hammer has been developed with the active participation and testing by many colleagues.
We especially thank from LHCb Juli\'an Garc\'\i a Pardi\~nas, Lucia Grillo, Donal Hill, Simone Meloni, Adam Morris, Patrick Owen, and Luke Scantlebury-Smead,
for their extensive feedback, discussions, questions, and beta testing during development. 
We similarly thank from Belle~II Kilian Lieret, Thomas Lueck, Felix Metzner, Markus Prim, and Maximilian Welsch.
We thank David Shih for discussions and comments on the manuscript.
Thanks are also due to all interested users on Belle, Belle~II, BaBar, LHCb, and CMS,
for many helpful discussions, questions, testing, and feedback. 
FB was supported by the DFG Emmy-Noether Grant No. BE 6075/1-1. 
SD was supported by the German Ministry of Research and Science (BMBF). 
ZL, MP and DR were supported in part by the Office of High Energy Physics of the 
U.S.\ Department of Energy under contract DE-AC02-05CH11231.
We thank the Aspen Center of Physics, supported by the NSF grant PHY-1607611, where parts of this work were completed.
This work also used resources of the National Energy Research Scientific Computing Center (NERSC), a U.S.\ Department of Energy Office of Science User Facility operated under Contract No. DE-AC02-05CH11231.
FB thanks Kim Scott, Bob Michaud and Julie Michaud-B for their hospitality, many good discussions and in general a great time in Houston, where part of this paper was written. 
\end{acknowledgements}

\appendix

\makeatletter
\renewcommand\thetocsection{\@Alph\c@section}
\makeatother

\section{Core elements of the Application Programming Interface}
\label{sec:API}

\begin{figure}[b]
	\includegraphics[width = \linewidth]{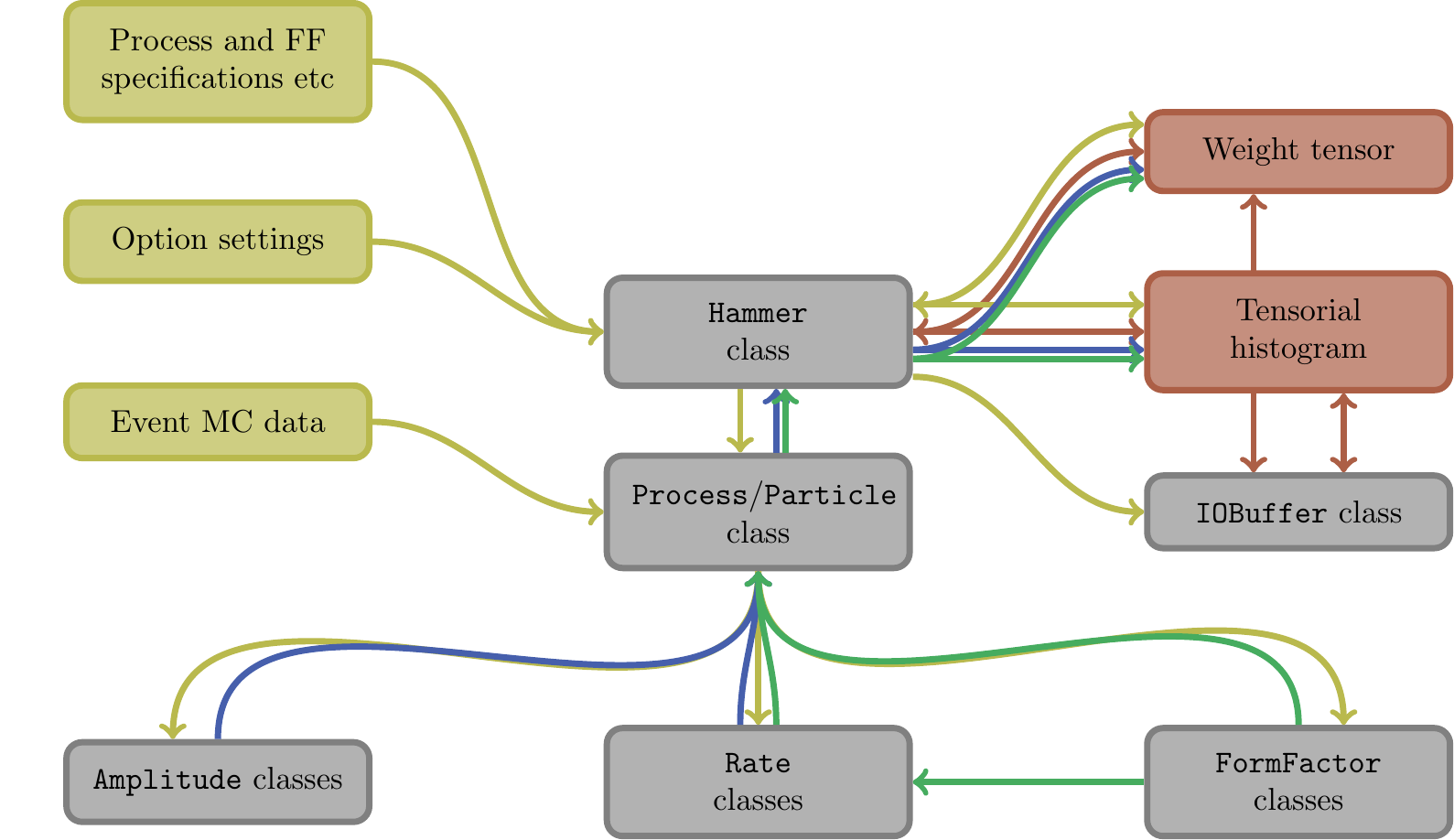}
	\caption{Schematic architecture of \hammer. The flow of user specified choices or event data is shown by yellow arrows. Blue (green) arrows denote the flow of calculational 
	information, in particular amplitude, weight or rate (form factor) tensors. 
	Red arrows highlight the flow of \hammer output, which may be saved or reloaded. Most internal \hammer classes are not shown in this schematic.}
	\label{fig:Scheme}
\end{figure}

The user interface of the \hammer library provides four main classes: the \textlst{Hammer} class itself; the \textlst{Process} and \textlst{Particle} classes, used to create events; 
and the \textlst{IOBuffer} class used for saving and loading precomputed objects. A schematic of the architecture of \hammer is shown in Fig.~\ref{fig:Scheme}.

In the following we describe various core parts of the \hammer Application Programming Interface (API), with many more details available in the code manual.
The library itself is implemented in \texttt{C++}, along with a \texttt{Python3} wrapper of the API; we will consider here the \texttt{C++} interface only.
This discussion is ordered by scope, rather than the typical code flow. Further details can be found in the \hammer manual~\cite{hammer_manual}.

\subsection{Building processes and events}

\begin{figure}[t]
\begin{center}
\begin{tikzpicture}[scale = 0.7, transform shape]
		\tikzstyle{level 1}=[sibling angle=50];
		\tikzstyle{level 2}=[sibling angle=50];
		\tikzstyle{level 3}=[sibling angle=50];
		\tikzstyle{every node}=[draw=black!80,line width=1.5pt,fill=white,circle,inner sep=2pt,minimum size = 12pt,align=center]
		\tikzstyle{nil}=[draw=none]
		\tikzstyle{edge from parent}=[draw=black!40, line width = 2pt]
		\node (P) {0} [grow cyclic,shape=circle,level distance=2cm,clockwise from=-65]
			child  { [grow cyclic,shape=circle,level distance=2cm,clockwise from=-15] node {1} edge from parent [draw = black!80, line width = 3pt]
				child  {node [nil] {4}}
				child  {[grow cyclic,shape=circle,level distance=2cm,clockwise from=-65]  node {5} edge from parent [draw = black!80, line width = 3pt]
					child {node [nil] {9}}
					child {node [nil] {10}}
				}
				child  {node [nil] {6}};
			}
			child  {[grow cyclic,shape=circle,level distance=2cm,clockwise from=-115] node {2} edge from parent [draw = black!80, line width = 3pt]
				child  {node [nil] {7}}
				child  {node [nil] {8}};
			}
			child  {node [nil] {3}};
\end{tikzpicture}
\caption{Example process tree for a decay cascade involving 10 particles (numbers), 4 vertices (circles) and 3 edges (dark lines).}
\label{fig:processtree}
\end{center}
\end{figure}
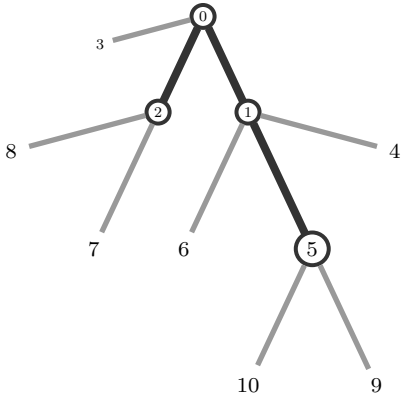

A typical decay cascade is contained in the library by the \textlst{Process} class; an event may contain multiple \textlst{Process} instances as e.g., is the case for a signal plus tag $B$-$\bar{B}$ pair. Each cascade may be simply represented in graphical terms as a `process tree', as shown in Fig.~\ref{fig:processtree}: 
Each particle in the cascade is assigned an index, and each decay vertex is represented as a map from a parent index, to the indices of all its daughters. 
\hammer assembles the process tree through two methods \textlst{Process::addParticle} and \textlst{Process::addVertex}. 
The former adds a \textlst{Particle} class object --  a momentum and a PDG code -- to a container of particles; the latter fills the map of each parent index to its daughters for each decay vertex.

In the case of Fig.~\ref{fig:processtree}, the first two vertices of the cascade may be built explicitly as follows:
\begin{lstcpp}
	Process proc;
	size_t idx0 = proc.addParticle(Particle{{E_0, px_0, py_0, pz_0}, pdg_0});
	size_t idx1 = proc.addParticle(Particle{{E_1, px_1, py_1, pz_1}, pdg_1});
	size_t idx2 = proc.addParticle(Particle{{E_2, px_2, py_2, pz_2}, pdg_2});
	size_t idx3 = proc.addParticle(Particle{{E_3, px_3, py_3, pz_3}, pdg_3});
	size_t idx7 = proc.addParticle(Particle{{E_7, px_7, py_7, pz_7}, pdg_7});
	size_t idx8 = proc.addParticle(Particle{{E_8, px_8, py_8, pz_8}, pdg_8});
	
	proc.addVertex(idx0, {idx1,idx2,idx3});
	proc.addVertex(idx2, {idx7,idx8});
\end{lstcpp}
and so on. Particles and vertices need not be added in order; helper functions are provided in the code examples for automatically parsing HepMC files.

\subsection{Specifications}
\label{app:spec}
The \hammer library contains an interpreter that maps a string representation of a vertex -- a vertex string -- to all possible charge conserving processes 
allowed by the specified particle names. 
The interpreter uses the syntax that particle names are parsed by a capital letter: the full list of names is provided in the manual. 
(For example the vertex string \textlst{"D*DPi"} is interpreted as all twelve possible $D^* \to D\pi$ vertices, 
while \textlst{"D*+DPi"} is interpreted as only the $D^{*+} \to D^+\pi^0$, $D^{*+} \to D^0\pi^+$, and (the heavily CKM suppressed) $D^{*+} \to \bar{D}^0\pi^+$ decay.)

The decay processes to be reweighed by \hammer are specified via \textlst{Hammer::includeDecay}, which takes a single vertex string or vector of vertex strings 
$\{V_1,V_2,\ldots, V_n\}$ as an argument, and may be invoked multiple times. 
Each \textlst{includeDecay} specification is \emph{inclusive} and permits any process tree whose full set of vertices contains all of $\{V_1,V_2,\ldots, V_n\}$. 
The boolean logic applied by \textlst{includeDecay} is \texttt{AND} between each vertex string element, and \texttt{OR} between separate invocations of \textlst{includeDecay}. 
For example
\begin{lstcpp}
	Hammer ham;
	ham.includeDecay({"BD*TauNu", "D*DGamma"});
	ham.includeDecay({"BDMuNu"});
\end{lstcpp}
means `Reweight a process that either contains vertices ($B \to D^*\tau\nu$ \textbf{and} $D^* \to D\gamma$) \textbf{or} the vertex ($B \to D \mu \nu$)'. 
Hence, e.g., $\bar{B}^0 \to (D^{*+} \to (D^+ \to K^+ \pi^+ \pi^-)\gamma)(\tau^- \to \ell^-\nu\nu)$ would be included. 
Recombination of radiative photons (produced during MC generation by \texttt{PHOTOS}) is handled automatically by the library, and need not be specified in \textlst{includeDecay} specifications. 
Processes may instead forbidden with the \textlst{Hammer::forbidDecay} method, 
whose specifications are \emph{exclusive} and forbids only process trees whose set of vertices $P$ \emph{equals} $\{V_1,V_2,\ldots, V_n\}$.

The \hammer library allows the user to specify multiple form factor `schemes' to be used in reweighting. 
A form factor scheme is a set of FF parameterization choices for each hadronic transition involving form factors, and is labelled by a `scheme name'. 
These schemes are set by the method \textlst{Hammer::addFFScheme}, which takes a scheme name plus a map from hadronic string representation to FF parametrization. 
The hadronic string follows the same syntax and uses the same particle symbols as for vertex strings. For example,
\begin{lstcpp}
	ham.addFFScheme("Scheme1", {{"BD", "BLPR"}, {"BD*", "BLPR"}};
	ham.addFFScheme("Scheme2", {{"BD", "BGL"}, {"BD*", "CLN"}});
\end{lstcpp}
declares two different FF schemes, choosing BLPR for both $B \to D$ and $B \to D^*$ form factors in \textlst{"Scheme1"}, and a mixture of schemes for \textlst{"Scheme2"}.
Separate histograms and event weights are generated for each scheme name. The list of available FF parametrizations are provided in Table~\ref{tab:knownampls}. 
The hadronic strings are charge sensitive, hence, e.g., \textlst{\{"B+D", "BLPR"\}} versus \textlst{\{"B0D", "CLN"\}} 
assigns two different FF parametrizations to charged and neutral $B \to D$ decays.
Specification of the form factor schemes used to generate the MC sample, i.e., the denominator or input form factors, 
must be specified in order for \hammer to be able to generate the reweighting tensors. These schemes are specified by the method \textlst{Hammer::setFFInputScheme}.

Units of the input MC sample may/should be specified via \textlst{Hammer::setUnits}, for instance \textlst{ham.setUnits("MeV")}. The default is GeV.

The \hammer library permits the user to declare particular vertices, in either the denominator or numerator amplitude, to be evaluated as pure phase space. 
This is achieved by the method \textlst{Hammer::addPurePSVertices}, which takes a set of string vertices as an argument, and an optional enum \textlst{WTerm} 
taking values \textlst{COMMON} (default), \textlst{NUMERATOR}, or \textlst{DENOMINATOR}. As an example
\begin{lstcpp}
	ham.addPurePSVertices({"TauMuNuNu", "D*+DPi"});
	ham.addPurePSVertices({"D*DGamma"}, WTerm::DENOMINATOR);
\end{lstcpp}
requests all $\tau \to \mu \nu \nu$ and $D^{*+}\to D\pi$ vertices in the numerator and all $D^* \to D\gamma$ vertices in the denominator, to be evaluated as phase space.
How these requests are enforced is subject to detailed rules explained in the manual. The library employs the pure phase space definition
\begin{equation}
	\label{eqn:PSdef}
	 \frac{1}{\prod_k |\{s_k\}|} \sum_{s_i, r_j} \big|\mathcal{M}_{s_1,\ldots, s_n; r_1,\ldots,r_m}\big|^2 = 1 \times (m^{6-2n})\,,
\end{equation}
where $s_i$ ($r_i$) are incoming (outgoing) quantum numbers, $|\{s_k\}|$ is the number of states of $s_k$, 
$m$ is the mass of the parent particle in the vertex, and $n$ the number of daughters.

Once all specifications are declared (include histograms, as below), containers are initialized by \textlst{Hammer::initRun()}. 
After invocation of \textlst{ham.initRun()}, manipulation of the FF default settings may be achieved via \textlst{setOptions}, which takes YAML\footnote{See \href{https://yaml.org}{yaml.org}.} format arguments. For instance,
\begin{lstcpp}
	ham.setOptions("BtoDBGL: {ChiTmB2: 0.01, ChiL: 0.002}");
\end{lstcpp}
changes the two BGL outer function parameters from their default settings. 
(Note that the YAML key for the relevant FF class has a \textlst{"to"} inserted in the hadronic transition, e.g., \textlst{"BtoDBGL"}, rather than \textlst{"BDBGL"}, 
to make it clear we are identifying settings for a particular class -- the $B \to D$ BGL class -- and not a process.) 

By default the library computes the total rate (or looks up a partial width) the first time each unique vertex is encountered in a run.
This behavior may be disabled, e.g., if the required integration is multidimensional and time consuming, via \textlst{ham.setOptions("ProcessCalc: \{Rates: false\}")}.

To permit full flexibility in FF settings, duplication of the same FF class is permitted and may be invoked by adding a token to a FF parametrization name in 
\textlst{addFFScheme}, separated by an underscore. 
For instance, one may declare
\begin{lstcpp}
	ham.addFFScheme("Scheme1", {{"B+D", "BGL_1"}, {"B0D", "BGL_2"},...});
	ham.addFFScheme("Scheme2", {{"BD", "BGL_3"},...});
\end{lstcpp}
This example allows independent manipulation of the BGL parameterization for each of the charged versus neutral modes in the same scheme, or between different schemes: 
After \textlst{initRun}, a succeeding \textlst{ham.setOptions("BtoDBGL_2:...")} would affect only the neutral $B$ parametrization in \textlst{"Scheme1"}.

Various other additional specification features are provided by the library, 
including e.g. specialization of Wilson coefficients to a particular global choice in the event tensor weight calculations. 
Specifications may also be declared through a card interface, as shown in \texttt{demo...card.cc} example programs provided with the source code.

\subsection{Histogramming}
Histograms of arbitrary dimensionality may be created by the \hammer library. 
In general, histogram bins contain event weight tensors (or direct products of them if there multiple processes in the event).

A histogram is declared by \textlst{Hammer::addHistogram}, which takes as arguments a name string and either: a vector of dimensions, 
a bool for under/overflow and a vector of ranges; or a vector of bin edges and a bool for under/overflow.
The method \textlst{addHistogram} does not create a single histogram, but rather a \emph{histogram set}: 
A separate histogram is created for each unique event cascade and in turn for each FF scheme name specified by \textlst{addFFScheme}. 
For instance
\begin{lstcpp}
	ham.addHistogram("q2VsEmu", {20, 15}, false,{{3.,12.},{0,2.5}});
	ham.addHistogram("q2VsEmu", {{3.,5.,9.,12.},{0,1,2.5}}, true);
\end{lstcpp}
The first declaration creates a \emph{histogram set} each with $20\times 15$ bins, no under/overflow, 
binned uniformly over the respective ranges $3$--$12$ and $0$--$2.5$ (in appropriate units).  
With reference to the above \textlst{addFFScheme} example, 
this histogram set contains one histogram for each combination of either \textlst{"Scheme1"} or \textlst{"Scheme2"} with each unique $B \to D$ decay cascade. 
The second declaration similarly creates a set of $3\times2$ histograms with non-uniform bins and additional under/overflow bins.

Filling of histograms for a specific event is performed by \textlst{Hammer::fillEventHistogram}, 
which takes the histogram name and the values of the observables corresponding to each histogram dimension. 
For example, \textlst{ham.fillEventHistogram("q2VsEmu", \{4., 0.5\})} fills the appropriate bin element for the \textlst{"q2VsEmu"} histograms belonging to the event being processed, 
and fills the relevant histograms for each declared FF scheme name.  
(Invocations of \textlst{fillEventHistogram} must occur before \textlst{Hammer::processEvent}, discussed in Sec.~\ref{sec:proc} below.)  
If \textlst{fillEventHistogram} is not invoked for a particular histogram for a particular event, the events tensor weight is not added to the histogram.
When the under/overflow bool is set to \textlst{false}, events outside the bin ranges are ignored by \textlst{fillEventHistogram}.

Computation of the weight-squared uncertainties is off by default. 
This may be enabled globally via the options setting \textlst{ham.setOptions("Histos: \{KeepErrors: true\}")}. 
However, for computational speed and/or memory efficiency, it may be instead enabled or disabled for individual histograms via \textlst{Hammer::keepErrorsInHistogram}, 
which takes the name of the histogram as an argument, and a bool. For instance
\begin{lstcpp}
	ham.keepErrorsInHistogram("q2VsEmu", true);
\end{lstcpp}
enables weight-squared computation for this particular histogram. This method should be invoked before \textlst{initRun}.

Various additional histogramming methods are provided by the library, that enable histogram compression, projection, and Wilson coefficient or FF specialization. 
These permit reduction of memory requirements or speed enhancements, and are detailed in the manual.

\subsection{Processing}
\label{sec:proc}

An event may contain multiple instances of \textlst{Process}, in order to account for the fact that a single event may feature, e.g., two $B$ decay processes.
The \textlst{Event} class is initialized by \textlst{Hammer::initEvent()}, 
which may take an optional initial event weight (this can also be set by \textlst{Hammer::setEventBaseWeight}).
\textlst{Process} instances are added by \textlst{Hammer::addProcess(proc)} which also returns a hash ID of the process.
If the process is not allowed according to the \textlst{includeDecay} or \textlst{forbidDecay} specifications, the returned hash ID is zero, and the process
is not added to the relevant \textlst{Event} containers.

Once a process is added, it is automatically initialized, which chiefly involves: calculating the signatures of each vertex in the decay cascade; 
identifying the various subamplitudes making up the cascade,
as well as relevant form factor parametrizations and vertex decay rates, for both the numerator/output and denominator/input; 
and calculating the total rate for the vertex (this is done only the first time each unique vertex is encountered, i.e., only once per run per unique vertex and per FF scheme). 
The amplitude tensors and form factors are not computed, however, until the invocation of \textlst{Hammer::processEvent}. 

Once all processes are added and relevant histograms (if any) have been denoted to be filled, 
the weights are actually computed and added to the histogram (if any) bins by invocation of \textlst{processEvent}. 
A pseudo-example on a single event with a set of processes might look like
\begin{lstcpp}
	ham.initEvent();
	bool isAllowed = false;
	//Create a set of Process, via addParticle and addVertex
	for(Process& proc: processes){
		auto procID = ham.addProcess(proc)
		if(procID != 0){
			//Calculate observables, fill histograms
			isAllowed = true;
		}
	}
	if(isAllowed){ ham.processEvent(); }
\end{lstcpp}
which might be emplaced in a larger loop over a set of events.

\subsection{Setting Wilson coefficients and form factors}
\label{sec:retrieve}. 

Once \textlst{processEvent} is completed, the event weight may be retrieved by \textlst{Hammer::getWeight}, that takes the FF scheme name. 
For instance \textlst{ham.getWeight("Scheme1")} computes the \emph{currently loaded event weight} for the \emph{currently specified WCs and FFs}. 
The latter may be set via \textlst{Hammer::setWilsonCoefficients} and \textlst{setFFEigenvectors}. 

The method \textlst{setWilsonCoefficients} takes a string that identifies which operator WCs are being set, and either a vector of the WC values or a map.
The default WC settings are the SM.  A typical example of the usage of this method is
\begin{lstcpp}
	 ham.setWilsonCoefficients("BtoCTauNu", 
	 		{{"S_qLlL", 1.}, {"T_qLlL",0.25}});
\end{lstcpp}
where the first argument specifies $b \to c \tau \nu$ four-Fermi WCs are being set, 
and the second argument is a \textlst{std::map<std::string, std::complex<double>>} of each WC to its desired value. The full list of WCs and their definitions is supplied in the manual. 
An optional third argument is the \textlst{WTerm} enum, that declares whether the evaluation should be applied to the numerator and/or denominator (numerator by default). 
As an alternative, one may instead pass as second argument a \textlst{std::vector<std::complex<double>>}, corresponding to the ordered basis
\begin{lstcpp}
	{"SM", "S_qLlL", "S_qRlL", "V_qLlL", "V_qRlL", 
		"T_qLlL", "S_qLlR", "S_qRlR", "V_qLlR", "V_qRlR", "T_qRlR"},
\end{lstcpp}
with the conventions for these WCs shown in Table~\ref{tab:NPc}. 
It is important to note that the \textlst{setWilsonCoefficients} method, when taking a \textlst{std::map}, produces \emph{incremental} settings changes. 
A subsequent invocation \textlst{ham.setWilsonCoefficients("BtoCTauNu", \{\{"S_qLlL", 0.5\}\})} will result in \textlst{S_qLlL} $= 0.5$ and \textlst{T_qLlL} $= 0.25$.
The method \textlst{resetWilsonCoefficients} resets the corresponding WCs to the default SM values. 

The FF eigenvectors are (re)set via the method \textlst{Hammer::setFFEigenvectors} (\textlst{resetWilsonCoefficients}) in a similar way, 
identifying the FF eigenvectors to be set via the FF class prefix such as \textlst{"BtoD"} and the parametrization name.  
A typical example of the usage of this method is
\begin{lstcpp}
	 ham.setFFEigenvectors("BtoD*", "BGLVar", 
	 	{{"delta_a1", 0.1}, {"delta_b1",-0.05}});
\end{lstcpp}
See the manual for definitions of currently implemented FF variational classes (typically denoted with a suffix \textlst{"Var"}).

\subsection{Retrieval}

Once all events or histograms have been processed (or reloaded from a file, see Sec.~\ref{sec:load}), 
the user may retrieve a specific histogram with the method \textlst{Hammer::getHistogram}, 
that takes a histogram name and a FF scheme name.  
NP choices must be specified first via \textlst{setWilsonCoefficients}, as must FF uncertainties via \textlst{setFFEigenvectors}. For example,
\begin{lstcpp}
	auto histo = ham.getHistogram("q2VsEmu","Scheme2");
\end{lstcpp}
would contract the bin weights with the specified NP Wilson coefficients (and FF eigenvectors, if any) for each histogram in the \textlst{"q2VsEmu"} histogram set 
with FF scheme \textlst{"Scheme2"}, and then combines them together into a single final histogram.
This contracted histogram output \textlst{histo} is a (row-major) flattened vector of \textlst{BinContents} structs. This struct has members \textlst{sumWi}, \textlst{sumWi2} and \textlst{n} 
for sum of weights, sum of squared weights and number of events in the bin, respectively.
(By contrast, the method \textlst{getHistograms} extracts all histograms of a specific name and scheme,
producing a map of event hash IDs to histogram for all available \textlst{"q2VsEmu"} histograms with the FF scheme \textlst{"Scheme2"}.)

Integrated rates or partial widths for a specific vertex may be retrieved via \textlst{Hammer::getRate}. 
The vertex is specified via either a vertex string, or the parent and daughter PDG codes, plus an FF scheme. 
(Partial widths are returned in the units specified by \textlst{Hammer::setUnits}; the default is GeV.) 
For example 
\begin{lstcpp}
	ham.getRate(511, {-413, -14, 13}, "Scheme2");
	ham.getRate("B0D*-MuNu", "Scheme2");
\end{lstcpp}
both return the partial width for the $B^0 \to D^{*-}\mu^+\nu$ vertex, using the form factor parameterization specified in \textlst{"Scheme2"}, 
and whatever WCs or FF uncertainties have been specified. 
(The \textlst{getRate} method is charge conjugate sensitive, so the vertex string must specify sufficient charges to make the vertex charge unique. 
For example, writing just \textlst{"B0D*MuNu"} would  correspond to not only $B^0 \to D^{*-}\mu^+\nu$, but also the very heavily suppressed $B^0 \to D^{+}\mu^-\bar{\nu}$.) 
The method \textlst{getDenominatorRate} similarly returns the partial width according to the specified denominator/input FF parametrization chosen in \textlst{setFFInputScheme}, 
and the denominator/input WCs or FF eigenvectors. 

\subsection{Multithreading}

The library has the ability to perform lock-free parallelization of the \textlst{getHistogram(s)} and \textlst{getWeight} evaluations.  
This requires use of the thread local methods \textlst{setWilsonCoefficientsLocal} and \textlst{setFFEigenvectorsLocal} to set the desired WC or FF uncertainties. 
These \textlst{...Local}  methods take the same syntax as \textlst{setWilsonCoefficients} and \textlst{setFFEigenvectors}, but with different behaviour: 
They do not set the values incrementally from the current settings, but always increment from the SM and zero FF uncertainties, respectively. 
Global values of the WCs or FF variations are unaffected by the \textlst{...Local} methods.

\subsection{Saving}

\hammer provides the ability to store header settings, generated event weights, histograms, and/or rates in binary buffers for later retrieval and reprocessing.
These buffers are built on the cross-platform serialization library \texttt{flatbuffers}\footnote{\href{https://google.github.io/flatbuffers/}{google.github.io/flatbuffers}.}: 
The buffer structs \textlst{Hammer::IOBuffer} and \textlst{Hammer::RootIOBuffer} permit writing/reading of \hammer internal objects using C++ binary files  and \texttt{ROOT} trees, respectively.

In order to save a buffer, an \textlst{ofstream} outfile must first be designated. For example, \textlst{ofstream outFile("./DemoSave.dat",ios::binary)}.
The methods \textlst{Hammer::saveRunHeader}, \textlst{saveEventWeights}, \textlst{saveRates}, \textlst{saveHistogram} may be used to save: specification settings 
(like \textlst{includeDecay} etc; the process weight(s) of the event currently loaded in memory (this should be invoked inside an event loop, after \textlst{processEvent}); 
the computed rates; and histograms. 
Each of these methods returns a \textlst{IOBuffer}, which can be stored as sequential records in the buffer via an ostream operator. 
For example, 
\begin{lstcpp}
	outFile << ham.saveRunHeader();
	outFile << ham.saveHistogram("q2VsEmu");
\end{lstcpp}
writes the declared run header, with all its settings, into an \textlst{IOBuffer} and passes it as a record into the buffer, and then does the same for the histogram \textlst{"q2VsEmu"}.
The record types are labelled by an \textlst{char} enum \textlst{Hammer::RecordType} with values 
\textlst{UNDEFINED = 'u'}, \textlst{HEADER = 'b'}, \textlst{EVENT = 'e'}, \textlst{HISTOGRAM = 'h'},  \textlst{HISTOGRAM_DEFINITION = 'd'}, and  \textlst{RATE = 'r'}. 
A histogram is always saved sequentially as a definition record then the histogram data record. 
The \textlst{saveHistogram} method may optionally take additional arguments -- such as an FF scheme name -- 
in order to save only part of an entire histogram set; see the manual for further details.

Saving a buffer in \texttt{ROOT} format is achieved by passing the \textlst{IOBuffer} output of the \textlst{save...} methods into a  \textlst{RootIOBuffer}, 
that may then be stored in a \texttt{ROOT} \textlst{TTree}. 
Explicit implementations of this functionality are provided in various \texttt{demo...root.cc} example programs.

\subsection{Reloading and merging}
\label{sec:load}
Buffer records may be loaded from a declared \textlst{ifstream} infile into an \textlst{IOBuffer} via an \textlst{istream} operator. For example,
\begin{lstcpp}
	ifstream inFile("./DemoSave.dat", ios::binary);
	Hammer::IOBuffer buf{Hammer::RecordType::UNDEFINED, 0ul, nullptr};
	inFile >> buf;
	ham.loadRunHeader(buf);
\end{lstcpp}
attempts to load the first buffer record as a run header (returning \textlst{false} if this record is of a different type). 

It is the responsibility of the user to curate the logic and order under which a buffer is saved and then read. 
For example, if a block of histograms have been saved before a set of rate records, then
\begin{lstcpp}
while(buf.kind != Hammer::RecordType::RATE) {
	if(buf.kind == Hammer::RecordType::HISTOGRAM) {
		ham.loadHistogram(buf);
	} 
	if(buf.kind == Hammer::RecordType::HISTOGRAM_DEFINITION){
		ham.loadHistogramDefinition(buf);
	}
	inFile >> buf;
}
\end{lstcpp}
would read through the buffer, with the method \textlst{Hammer::loadHistogram} loading all the histograms, and \textlst{Hammer::loadHistogramDefinition} 
all the histogram definitions, that are found before reaching the block of saved rates.
The method \textlst{loadRates} behaves similarly to \textlst{loadHistogram}. 

Once an object is loaded, it behaves just as the originally computed instance. 
So one may invoke \textlst{getHistogram} for a reloaded histogram as described in Sec.~\ref{sec:retrieve}. 

Event weights can be reloaded via \textlst{loadEventWeights}. This permits recreating the original event loop provided \textlst{initEvent} and \textlst{processEvent} are called appropriately. For example, on a block of saved event records
\begin{lstcpp}
while(buf.kind == Hammer::RecordType::EVENT) {
	ham.initEvent();
	ham.loadEventWeights(buf);
	double q2 = ...; //Calculate q^2 from known kinematic event information 
	ham.fillEventHistogram("Q2", {q2});
	ham.processEvent();
	inFile >> buf;
}
\end{lstcpp}			
would permit reprocessing of saved event weights into a newly created \textlst{"Q2"} histogram.

Loading a buffer in \texttt{ROOT} format is achieved by reading the \textlst{RootIOBuffer} stored in a \textlst{TTree} into an \textlst{IOBuffer} that can be passed to the \textlst{load...} 
methods. Explicit implementations of this functionality are provided in various \texttt{demo...root.cc} example programs.

In order to permit parallelization of initialization runs, the \textlst{load...} methods accept an additional bool, 
to specify whether to \emph{merge} the buffer contents with existing objects in memory (\textlst{true}), or overwrite them (\textlst{false}, default). 
Merging of histograms occurs if two histograms are loaded with a matching name. 
This merging is \emph{additive} for histograms in each histogram set with the same FF scheme and hash IDs, 
and otherwise results in the new unique histograms being \emph{appended} to the existing histogram set. 
(If one wishes instead to overwrite a histogram one may instead first invoke
\textlst{removeHistogram}, and then reload the desired components of the histogram set.)

The methods \textlst{loadEventWeights} and \textlst{loadRates} behave similarly. For weights (rates) with matching hash IDs, 
merging permits appending of process weights (rates) computed with new form factor schemes to the process weights (rates). 
Finally, \textlst{loadRunHeader} permits merging of two sets of header specifications into their union. More details are provided in the manual.

\providecommand{\href}[2]{#2}\begingroup\raggedright\endgroup

\end{document}